\documentclass[a4paper]{article}

\usepackage{graphicx}
\usepackage{epsfig}
\usepackage[fleqn]{amsmath}
\usepackage[psamsfonts]{amssymb}
\usepackage[varg]{txfonts}

\usepackage{lineno}

\title{{\bf A Study on Optimal Beam Patterns for Single User Massive MIMO Transmissions}}

\author{Maki Arai, Kei Sakaguchi, Kiyomichi Araki \\ 
	E-mail: arai@mobile.ee.titech.ac.jp}
\date{}

\newcommand{\argmax}{\mathop{\rm arg~max}\limits}

\newcommand{\mrm}{\mathrm}
\newcommand{\mbf}{\mathbf}
\newcommand{\mbb}{\mathbb}

\newcommand*\patchAmsMathEnvironmentForLineno[1]{%
  \expandafter\let\csname old#1\expandafter\endcsname\csname #1\endcsname
  \expandafter\let\csname oldend#1\expandafter\endcsname\csname end#1\endcsname
  \renewenvironment{#1}%
     {\linenomath\csname old#1\endcsname}%
     {\csname oldend#1\endcsname\endlinenomath}}%
\newcommand*\patchBothAmsMathEnvironmentsForLineno[1]{%
  \patchAmsMathEnvironmentForLineno{#1}%
  \patchAmsMathEnvironmentForLineno{#1*}}%
\AtBeginDocument{%
\patchBothAmsMathEnvironmentsForLineno{equation}%
\patchBothAmsMathEnvironmentsForLineno{align}%
\patchBothAmsMathEnvironmentsForLineno{flalign}%
\patchBothAmsMathEnvironmentsForLineno{alignat}%
\patchBothAmsMathEnvironmentsForLineno{gather}%
\patchBothAmsMathEnvironmentsForLineno{multline}%
}

\begin{document}
\maketitle
\section*{Summary}
This paper proposes optimal beam patterns of analog beamforming 
for SU (Single User) massive MIMO (Multi-Input Multi-Output) transmission systems. 
For hybrid beamforming in SU massive MIMO systems,
there are several design parameters
such as beam patterns, the number of beams (streams), 
the shape of array antennas, and so on. 
In conventional hybrid beamforming, 
rectangular patch array antennas implemented on a planar surface 
with linear phase shift beam patterns have been used widely. 
However, 
it remains unclear whether existing configurations are optimal or not.
Therefore, we propose a method using OBPB (Optimal Beam Projection Beamforming) 
for designing configuration parameters of the hybrid beamforming. 
By using the method, the optimal beam patterns are derived first,
and are projected on the assumed surface to calculate the achievable number of streams 
and 
the resulting channel capacity. 
The results indicate OBPB with a spherical surface yields at least 3.5 times higher channel capacity than conventional configurations.

{\bf keywords:}
massive MIMO, beamforming, beam pattern, directivity, spherical mode expansion, capacity maximization.

\section{Introduction}
Future wireless communication systems must offer larger channel capacity
because of the popularization of wireless devices 
such as smartphones and tablets. 
To increase the channel capacity, massive MIMO (Multi-Input Multi-Output) technology 
using a large number of 
antenna elements, at least at the BS (Base Station), is important \cite{Lu,Ji}, 
and the technology is expected to be deployed in 5G cellular networks and beyond. 
In the case of MU (Multi-User) massive MIMO, 
the channel (system) capacity increases in proportion to the number of antenna elements 
if UEs (User Equipment) are well separated in space \cite{Nam}. 
However, in the case of SU (Single User) massive MIMO, 
this is not true and the number of streams for spatial multiplexing is limited due to both smaller antenna surface at the UE and the increased antenna correlation at the BS.

Hybrid beamforming, which combines analog beamforming and digital pre/post-processing, is
a reasonable way 
to realize massive MIMO systems in low cost and low power consumption \cite{Sohrabi,Obara,Xiao,Nsenga}. 
In conventional hybrid beamforming, 
rectangular patch array antennas implemented on a planar surface with linear phase shift beam patterns have been used widely \cite{Ng,Swindlehurst}. 
In the case of MU massive MIMO with UEs well separated in the space, 
it works well by just steering main beams to locations (angles seen from the BS) of the UEs. 
However, in the case of SU massive MIMO, 
it will not work well 
since steering main beams to the single UE is just increasing antenna correlation at the BS. 
Therefore, there are remaining research issues in the case of SU massive MIMO 
in terms of beam patterns, the number of beams (streams), 
the shape of 
an array antenna, and so on.

In this paper, we introduce a method using OBPB (Optimal Beam Projection Beamforming) 
proposed in \cite{Arai_IEICE} 
for designing antenna configuration parameters of the SU massive MIMO. 
The goal of the study is 
to maximize channel capacity 
by increasing the effective number of beams (streams) 
in a given propagation channel with special beam patterns and 
shapes of antenna designed by the OBPB. 
Different from the conventional design methods, 
the OBPB derives optimal beam patterns to be matched with the propagation channel first, 
and projects the optimal beam patterns to the assumed antenna surface 
such as sphere to synthesize conditional beam patterns. 
In the method that we proposed, 
the optimal solutions of the transmit and receive sides are derived 
by sequential calculations on a computer. 
Thus, there is no information exchanged between BS and UE. 
The necessary information to derive the solutions is only the joint angular profile
to make the corresponding antenna radiation patterns for usage.
Thus, only measurements and feedbacks to determine the joint angular profile
are required without any other information exchanges to calculate
the optimal patterns even in real operation. 
Since OBPB utilizes as much space of the antenna surface as possible, the
larger number of orthogonal beams (streams) can be created. 
Table\ \ref{tbl:AnalogBeamforming} compares the antenna configuration and metrics of OBPB
with those of conventional methods for the analog beamforming.
From the analysis, 
it is found that the channel capacity realized by the OBPB 
with a spherical surface 
approaches the optimal capacity and 
is 3.5 times or larger than that of the conventional configurations. 

This paper is organized as follows. 
In Sect.\ 2, system model of SU massive MIMO and antenna configuration of the conventional hybrid beamforming are described. 
In Sect.\ 3, the proposed method using OBPB is introduced to design conditionally optimal antenna configuration parameters. 
Section\ 4 designs conditionally optimal beam patterns in a given environment and calculate the achievable number of streams and the resulting channel capacity with deep discussions about the results. 
Finally, Sect.\ 5 concludes this paper.

\begin{table*}[tb]
	\begin{center}
	\caption{Analog beamforming configuration.}
	\label{tbl:AnalogBeamforming}
{\footnotesize
		\begin{tabular}{|c||c|c|}
		\hline
		& Conventional & Proposed \\
		\hline
		\hline
		Shape & Planar & Spherical, Planar  \\
		\hline
		Category & Full-array, Sub-array & Hemisphere, 1/32-sphere, Plane  \\
		\hline
		Beamforming & Linear Phase Shift Beamforming (LPSB) 
		& Optimal Beam Projection Beamforming (OBPB) \cite{Arai_IEICE} \\	
		\hline	
		Antenna type & Patch array & Continuous surface \\
		\hline
		Beam selection metric & Received power, Determinant & Determinant \\
		\hline
		Rank adaptation metric & Capacity & Capacity \\
		\hline
		\end{tabular}
}
	\end{center}
\end{table*}

\section{Conventional hybrid beamforming for SU-massive MIMO system}
There are two categories of configurations, such as full-array and sub-array
in the conventional hybrid beamforming for the SU-massive MIMO.
The antenna elements are considered as 
the rectangular patch array antennas implemented on a planar surface 
with LPSB (Linear Phase Shift Beamforming).
The beam patterns are selected to maximize 
the received power of each stream 
or to maximize 
the determinant of a channel correlation matrix by using a combinatorial search.

\subsection{SU-massive MIMO system model}
\begin{figure}[tb]
    \centering
    \includegraphics[width=0.42\textwidth]{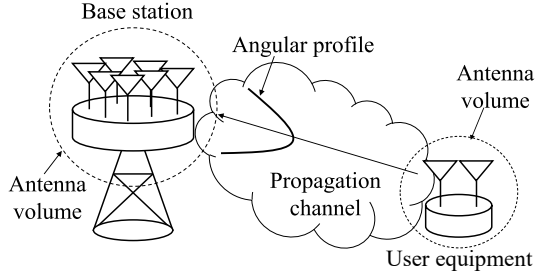}
    \caption{SU-massive MIMO system model.}
    \label{fig:SystemModel}
\end{figure}

\begin{figure}[tb]
    \centering
    \includegraphics[width=0.45\textwidth]{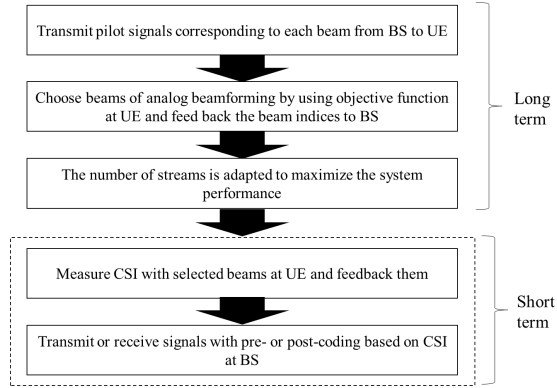}
    \caption{Hybrid beamforming procedure of massive MIMO system.}
    \label{fig:MassiveMIMOFlow}
\end{figure}

The massive MIMO system is achieved 
by using dozens, hundreds or more antenna elements 
at least at the BS
to improve channel (system) capacity as shown in Fig.\ \ref{fig:SystemModel}.
In the massive MIMO system, 
combining analog beamforming and digital 
pre/post-processing is 
a reasonable way 
for low cost and low power consumption. 
It is called as hybrid beamforming and 
its procedure consists of long-term and short-term operations 
as shown in Fig.\ \ref{fig:MassiveMIMOFlow}. 

The receive uplink signals of $M$ streams in the SU-massive MIMO system at time $t$ is defined as 
\begin{align}
	\mbf{y}_\mrm{BS}(t) 
	&= \mbf{W}_\mrm{d}^\mrm{T}(t) \mbf{W}_\mrm{a}^\mrm{T} \mbf{H}_0(t) \mbf{s}_\mrm{UE}(t) + \mbf{n}_\mrm{BS}(t) \nonumber \\
	&= \mbf{W}_\mrm{d}^\mrm{T}(t) \mbf{H}(t) \mbf{s}_\mrm{UE}(t) + \mbf{n}_\mrm{BS}(t) \\
	\mbf{H}(t) & = \mbf{W}_\mrm{a}^\mrm{T} \mbf{H}_0(t),
\end{align}
where 
$\mbf{W}_\mrm{d}(t) \! \in \! \mbb{C}^{M \times M}$ is a digital 
pre/post-processing weight matrix, 
$\mbf{W}_\mrm{a} \! \in \! \mbb{C}^{N_\mrm{BS} \times M}$ is an analog beamforming weight matrix,
$\mbf{s}_\mrm{UE} \! \in \! \mbb{C}^{N_\mrm{UE} \times 1}$ 
is a vector of transmit signal
and $\mbf{n}_\mrm{BS} \! \in \! \mbb{C}^{M \times 1}$ 
is a noise vector. 
$\mbf{H}_0(t)$ is a channel matrix
and its component is defined by a channel response between the $i$-th BS antenna and the $j$-th UE antenna $h_{0,ij}$. 
$\mbf{H}(t)$ is a channel matrix including the analog beamforming weight matrix. 
$M$ is the number of streams defined as $M \leq \min \{N_\mrm{BS}, N_\mrm{UE} \}$
and $N_\mrm{BS}, N_\mrm{UE}$ are the numbers of BS and UE antennas respectively.
In this paper, 
it is assumed that 
the transmit power is divided equally for all streams. 
Under the equally distributed power condition, 
the instantaneous channel capacity is derived as follows. 
\begin{align}
\label{eq:C}
	C(t) & = \log_2 \mrm{det} \left( \mbf{I}_M + \mbf{W}_\mrm{d}^\mrm{T}(t)
	\mbf{H}(t) \mbf{H}^\mrm{H}(t)  \mbf{W}_\mrm{d}^\mrm{*}(t)
	\frac{P}{M P_\mrm{n}}  \right), 
\end{align}
where $\mbf{I}_M$ is 
an $M \times M$ unity matrix, 
$P$ is the transmit power and $P_\mrm{n}$ is 
the noise power.
When an SVD (Singular Value Decomposition) is considered, 
the digital signal processing weight matrix is unitary. 
Thus, the channel capacity is expressed as
\begin{align}
\label{eq:C_2}
	C(t) & = \log_2 \mrm{det} \left( \mbf{I}_M + 
	\mbf{H}(t) \mbf{H}^\mrm{H}(t) 
	\frac{P}{M P_\mrm{n}}  \right).
\end{align}


\subsection{Antenna configuration for hybrid beamforming}
\begin{figure}[tb]
    \centering
    \includegraphics[width=0.42\textwidth]{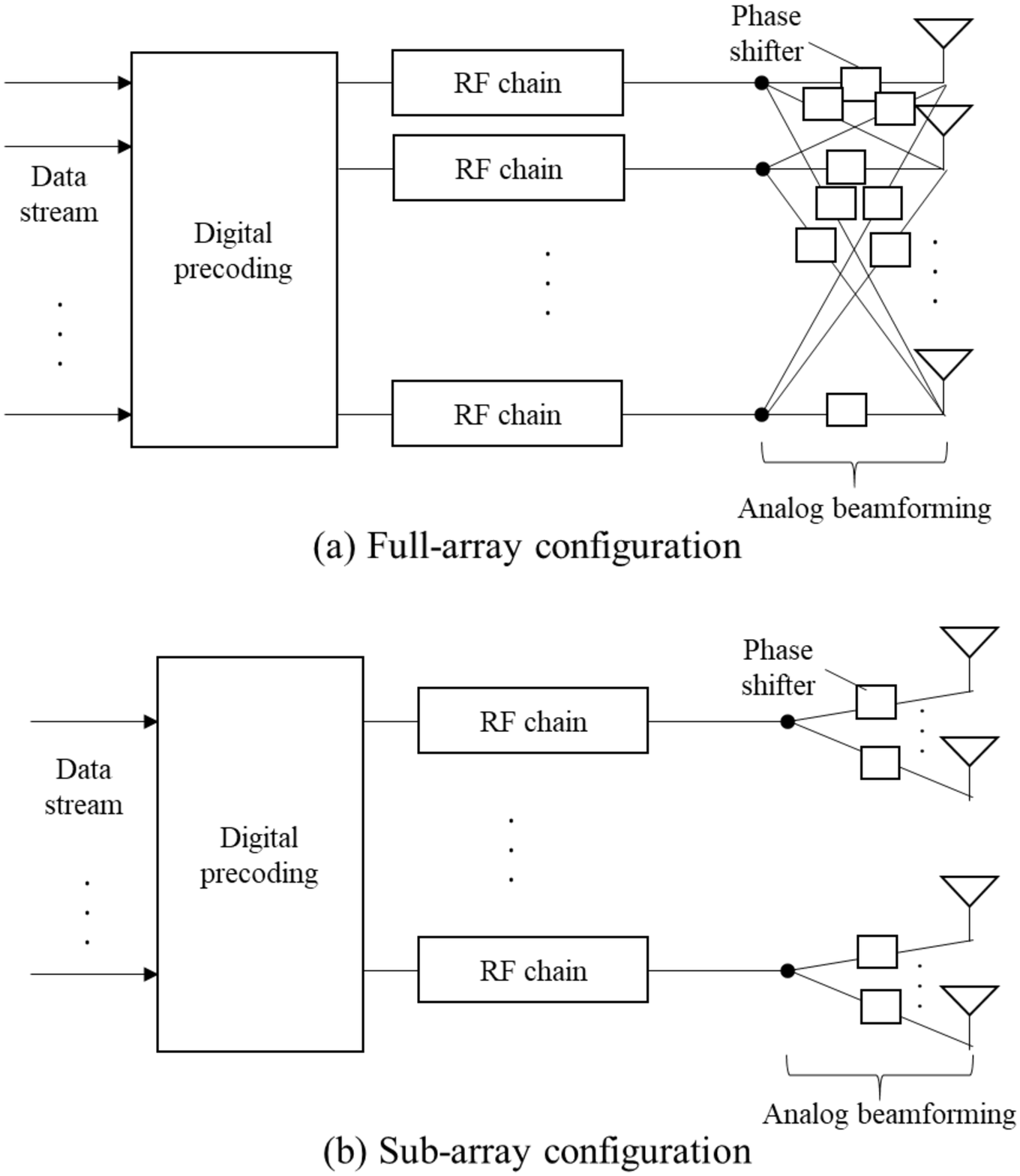}
    \caption{Hybrid beamforming antenna configuration.}
    \label{fig:Array}
\end{figure}

There are mainly two types of 
massive MIMO antenna configurations, 
i.e. the full-array and sub-array as shown in Fig.\ \ref{fig:Array}.
In the configurations, phase shifters are used to achieve analog beamforming weights. 
In the case of the full-array configuration, 
each RF chain is connected to all antenna elements.
The weight vector between the $m$-th RF chain and $N_\mrm{BS}$ antenna elements is given by
\begin{align}
	\mbf{w}_{\mrm{a},m} = [w_{\mrm{a},1m}, \cdots, w_{\mrm{a},N_\mrm{BS}m}]^\mrm{T}. 
\end{align}
The analog beamforming weight matrix for the full-array configuration is expressed as
\begin{align}
	\mbf{W}_\mrm{a} = [\mbf{w}_{\mrm{a},1}, \cdots, \mbf{w}_{\mrm{a},M}]. 
\end{align}

On the other hand, in the sub-array configuration, the antennas are divided 
into several groups
and each RF chain is connected to  a sub-array group with $N_\mrm{BS,sub}$ antenna elements, 
defined as $N_\mrm{BS,sub}=N_\mrm{BS}/M=N_\mrm{BS,sub,V}\times N_\mrm{BS,sub,H}$ sub-array antenna elements 
in order to simplify the feeding circuit.
The weight vector between the $m$-th RF chain and $N_\mrm{BS,sub}$ antenna elements is given by
\begin{align}
	\mbf{w}_{\mrm{a},m} = [w_{\mrm{a},1m}, \cdots, w_{\mrm{a},N_\mrm{BS,sub}m}]^\mrm{T}. 
\end{align}
The analog beamforming weight matrix for the sub-array configuration is expressed as
\begin{align}
	\mbf{W}_\mrm{a} = 
	\left[ 
		\begin{array}{cccc}
		\mbf{w}_{\mrm{a},1} & \mbf{0} & \cdots & \mbf{0} \\
		\mbf{0} & \mbf{w}_{\mrm{a},2} & & \vdots \\
		\vdots & & \ddots & \mbf{0} \\
		\mbf{0} & \cdots & \mbf{0} & \mbf{w}_{\mrm{a},M} \\
		\end{array} 
	\right]. 
\end{align}

\subsection{Analog beamforming weight matrix by using LPSB}
\begin{figure}[tb]
    \centering
    \includegraphics[width=0.34\textwidth]{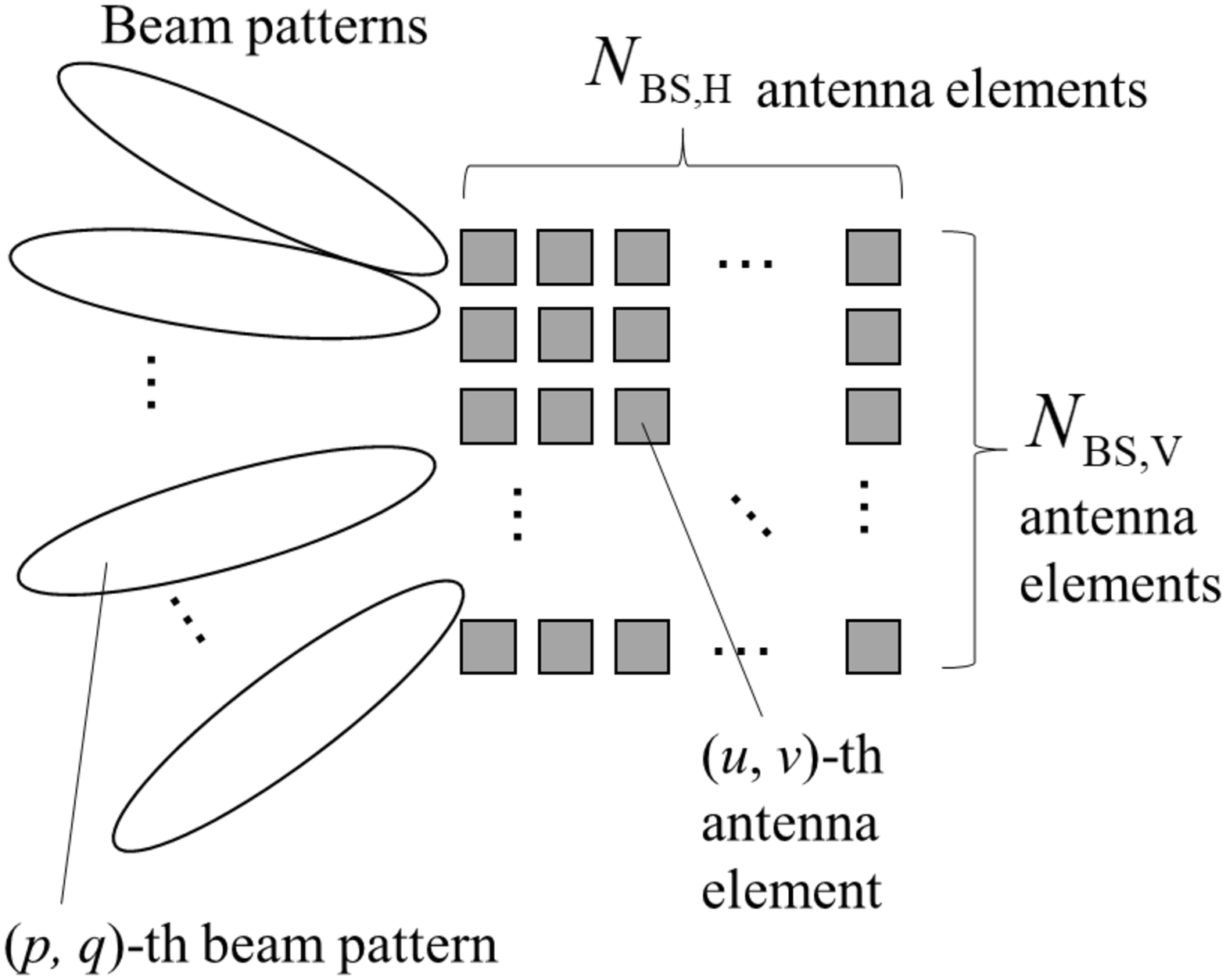}
    \caption{2-dimensional rectangular full-array configuration.}
    \label{fig:FullArrayAntennas}
    \centering
    \includegraphics[width=0.4\textwidth]{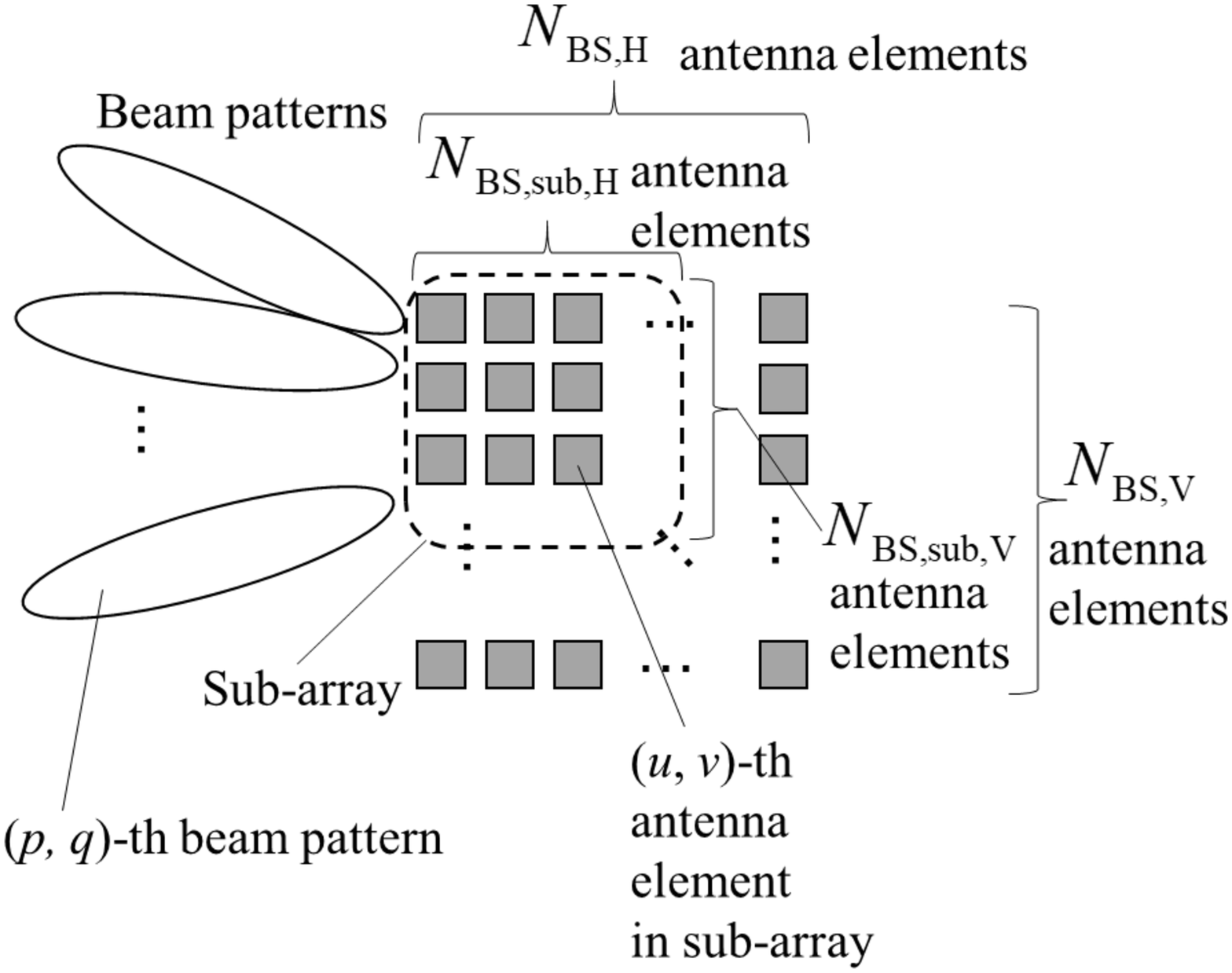}
    \caption{2-dimensional rectangular sub-array configuration.}
    \label{fig:SubArrayAntennas}
\end{figure}

When the linear phase shift beam patterns are used for the analog beamforming, 
the weight matrix is expressed based on a DFT (Discrete Fourier Transform) matrix 
$\mbf{D} \! \in \! \mbb{C}^{N_\mrm{BS} \times a^2 N_\mrm{BS}}$,  
where $a$ is 
a beam interval coefficient determined by an integer. 
For example, it is  
assumed in Fig.\ \ref{fig:FullArrayAntennas}
that 2-dimensional $N_\mrm{BS,V} \times N_\mrm{BS,H}$ rectangular array is used for 
the full-array.
The weight component for the $n=(u,v)$-th antenna element with the phase shifter
corresponding to the $m=(p,q)$-th beam pattern is expressed as follows.
\begin{align}
	d_{nm}
	= d_{up} d_{vq}
	& = \frac{1}{\sqrt{N_\mrm{BS}}}
	\exp \left(\frac{-j2\pi (u-1)(p-1)}{a N_\mrm{BS,V}} \right) \nonumber \\
	& \cdot \exp \left(\frac{-j2\pi (v-1)(q-1)}{a N_\mrm{BS,H}} \right),
\end{align}
where $u,v$ are the row indices of the DFT matrix 
in $\{ 1, \cdots, N_\mrm{BS,V} \}$ and $\{ 1, \cdots, N_\mrm{BS,H} \}$ respectively.
$p,q$ are the column indices of the DFT matrix
in $\{ 1, \cdots, a N_\mrm{BS,V} \}$ and $\{ 1, \cdots, a N_\mrm{BS,H} \}$.
When the value $a$ becomes large, the main lobes of the beams become near. 

In the case of the sub-array as shown in Fig.\ \ref{fig:SubArrayAntennas}, 
the weight component for the $n=(u,v)$-th antenna element with the phase shifter
corresponding to the $m=(p,q)$-th beam pattern is also expressed as 
\begin{align}
	d_{nm}
	= d_{up} d_{vq}
	& = \frac{1}{\sqrt{N_\mrm{BS,sub}}}
	\exp \left(\frac{-j2\pi (u-1)(p-1)}{a N_\mrm{BS,sub,V}} \right) \nonumber \\
	& \cdot \exp \left(\frac{-j2\pi (v-1)(q-1)}{a N_\mrm{BS,sub,H}} \right),
\end{align}
where the indices $u,v$ are integers  
in $\{ 1, \cdots, N_\mrm{BS,sub,V} \}$ and $\{ 1, \cdots, N_\mrm{BS,sub,H} \}$ 
corresponding to the row indices of the DFT matrix.
and the indices $p,q$ are integers  
in $\{ 1, \cdots, a N_\mrm{BS,sub,V} \}$ and $\{ 1, \cdots, a N_\mrm{BS,sub,H} \}$ 
corresponding to the column indices of the DFT matrix.

\subsection{Beam selection using received power of each stream}
We consider two 
methods for 
selecting analog beamforming weights.
First, the analog beamforming weight vectors are chosen to maximize the received power of each stream.
Next, 
the weight is selected to be descending order 
of the achievable received power of the stream.
\begin{align}
	\mbf{w}_m & = \argmax _{\mbf{d}_{m'} \in \mbf{D} \backslash \mbf{W}_{m-1}} 
	\mrm{E}[ \mbf{d}_{m'}^\mrm{T} \mbf{H}(t) \mbf{H}^\mrm{H} (t) \mbf{d}_{m'}^\mrm{*} ] \\
	\mbf{W}_{m-1} & = [\mbf{w}_1, \cdots, \mbf{w}_{m-1}], 
\end{align}
where $\mbf{d}_{m'}$ is the $m'$-th column vector of the matrix $\mbf{D}$. 
This method is simple because the only BS or UE side's information is needed for the calculation. 
However, the channel capacity degrades
due to 
high correlation between the selected beams
made by analog beamforming
in this method.

\subsection{Beam selection using determinant of channel correlation matrix}
Since the analog weight matrix is deterministic, 
the optimal values are derived by using
not 
the instantaneous channel capacity but the average channel capacity. 
When $\frac{P_\mrm{t}}{M P_\mrm{n}}$ is sufficiently large, 
the average channel capacity is given by 
\begin{align}
\label{eq:AverageC}
	\bar{C}& = \mrm{E} [C(t)] \nonumber \\
	& = \log_2 \mrm{det} \left( \mrm{E} \left[\mbf{I}_M + 
	\mbf{H}(t) \mbf{H}^\mrm{H}(t) 
	\right] \frac{P}{M P_\mrm{n}}  \right) \nonumber \\
	& \approx \log_2 \mrm{det} \left( \mrm{E} \left[
	\mbf{H}(t) \mbf{H}^\mrm{H}(t) 
	\right] \frac{P_\mrm{t}}{M P_\mrm{n}}  \right) \nonumber \\
	& \propto \mrm{det} \mrm{E} \left[ \mbf{H}(t) \mbf{H}^\mrm{H}(t) \right], 
\end{align}
Therefore, to maximize the average channel capacity 
is equivalent to maximize the determinant of 
the channel correlation matrix.
\begin{align}
	\mbf{w}_m & = \argmax_{\mbf{d}_{m'} \in \mbf{D} \backslash \mbf{W}_{m-1}} 
	\mrm{det} \left(
	\mrm{E} [\tilde{\mbf{W}}_{mm'}^\mrm{T} \mbf{H}(t) \mbf{H}^\mrm{H} (t) \tilde{\mbf{W}}_{mm'}^\mrm{*} ] \right) \\
	\mbf{W}_{m-1} & = [\mbf{w}_1, \cdots, \mbf{w}_{m-1}] \\
	\tilde{\mbf{W}}_{mm'} & = [\mbf{w}_1, \cdots, \mbf{w}_{m-1}, \mbf{d}_{m'}].
\end{align}
By using this method, the weights can be derived considering both 
the maximization of the beam gain and the reduction of 
correlation between the selected beams.
Therefore, the channel capacity can be improved compared to the first method using the received power.

\section{Analog beamforming by using OBPB}
OBPB is used to derive the effective number of beams (streams) and beam patterns 
and described by using the system model based on 
an SME (Spherical Mode Expansion) \cite{Hansen}. 
By using OBPB, 
optimal beam patterns are derived first to maximize the average channel capacity 
with the given propagation channel. 
After that, semi-optimal beam patterns are calculated under a given condition of antenna surface.
In this section, we introduce how to derive the semi-optimal beam patterns  
by projecting the optimal beam patterns to the assumed antenna surface 
and synthesizing the conditional beam patterns radiated from the surface.

\subsection{SU-massive MIMO system model with SME}
\begin{figure}[tb]
    \centering
    \includegraphics[width=0.3\textwidth]{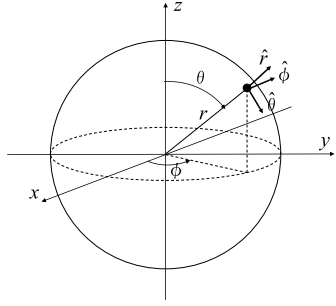}
    \caption{Spherical coordinate.}
    \label{fig:SphericalCoordinate}
\end{figure}

The channel matrix is expressed by using BS and UE antenna directivities as follows. 
\begin{align}
	& \mbf{H}(t) \nonumber \\
	& = \int_{\psi_\mrm{BS}} \! \int_{\psi_\mrm{UE}} \!
	\vec{\mbf{g}}_\mrm{BS}(\psi_\mrm{BS}) 
	\vec{\vec{h}}(\psi_\mrm{UE},\psi_\mrm{BS},t)
	\vec{\mbf{g}}_\mrm{UE}(\psi_\mrm{UE}) 
	\mrm{d}\psi_\mrm{UE} \mrm{d}\psi_\mrm{BS} \nonumber \\
	& = \mbf{Q}_{\mrm{BS},M}^\mrm{T} 
	\int_{\psi_\mrm{BS}} \! \int_{\psi_\mrm{UE}} \!
	\vec{\mbf{k}}_\mrm{BS}(\psi_\mrm{BS}) 
	\vec{\vec{h}}(\psi_\mrm{UE},\psi_\mrm{BS},t) \nonumber \\
	& \ \ \ \ \ \  \vec{\mbf{k}}_\mrm{UE}(\psi_\mrm{UE}) 
	\mrm{d}\psi_\mrm{UE} \mrm{d}\psi_\mrm{BS}
	\mbf{Q}_{\mrm{UE},M}^* \\
	& \vec{\mbf{g}}_\mrm{BS}(\psi_\mrm{BS})
	= \mbf{Q}_{\mrm{BS},M}^\mrm{T} \vec{\mbf{k}}_\mrm{BS} (\psi_\mrm{BS}) \\
	& \vec{\mbf{g}}_\mrm{UE}(\psi_\mrm{UE}) 
	= \mbf{Q}_{\mrm{UE},M}^\mrm{T} \vec{\mbf{k}}_\mrm{UE} (\psi_\mrm{UE}),	
\end{align}
where the
departure or arrival angles at BS and UE are $\psi_\mrm{BS}=(\theta_\mrm{BS},\phi_\mrm{BS})$ in a spherical coordinate shown in Fig. \ref{fig:SphericalCoordinate}. 
And $\psi_\mrm{UE}=(\theta_\mrm{UE},\phi_\mrm{UE})$. 
$\mbf{Q}_{\mrm{BS},M} \! \in \! \mbb{C}^{J_\mrm{BS} \times M},
 \mbf{Q}_{\mrm{UE},M} \! \in \! \mbb{C}^{J_\mrm{UE} \times M}$ are matrices of spherical mode coefficients
which determine the beam patterns for $M$ streams.
$J_\mrm{BS}$ and $J_\mrm{UE}$ 
are numbers of the spherical modes. 
$\vec{\mbf{k}}_\mrm{BS}(\psi_\mrm{BS}) \! \in \! \mbb{C}^{J_\mrm{BS} \times 1}, 
\vec{\mbf{k}}_\mrm{UE}(\psi_\mrm{UE})  \! \in \! \mbb{C}^{J_\mrm{UE} \times 1}$
are vectors of far-field pattern functions which are canonical solutions of Helmholtz equation.

\begin{figure}[tb]
    \centering
    \includegraphics[width=0.49\textwidth]{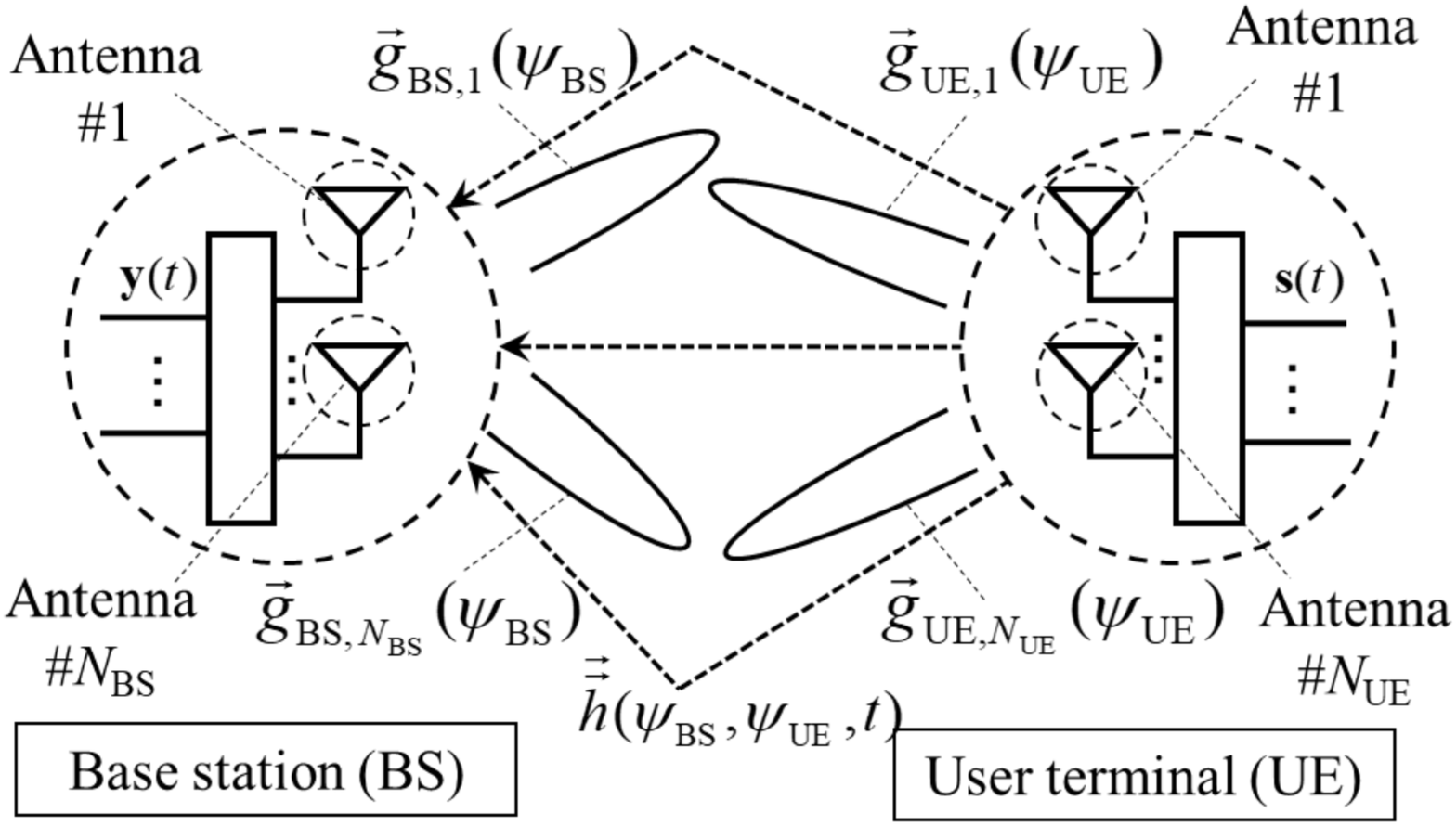}
    \caption{MIMO system model with SME.}
    \label{fig:MIMOwithSME}
\end{figure}

From Eq.\ (\ref{eq:AverageC}), the maximization of the average channel capacity is 
equal to that of the determinant of the channel correlation matrix.
When SME is used, the channel correlation matrix which 
transmits and receives $M$ streams 
is expressed as follows.
\begin{align}
\label{eq:R_BS_h}
	\mbf{R}_{\mrm{BS,h}}
	&= \mrm{E} [\mbf{H} (t) \mbf{H}^\mrm{H} (t) ] \nonumber \\
    &= \int_{\psi_{\mrm{BS}}} \int_{\psi_{\mrm{UE}}}
    \vec{\mbf{g}}_{\mrm{BS}}(\psi_{\mrm{BS}}) \cdot 
    \Big( \vec{\mbf{g}}^{\mrm{T}}_{\mrm{UE}}(\psi_{\mrm{UE}}) 
    \cdot \vec{\vec{P}}_{\mrm{h}}(\psi_{\mrm{BS}},\psi_{\mrm{UE}}) \nonumber \\
	& \cdot
    \vec{\mbf{g}}^{*}_{\mrm{UE}}(\psi_{\mrm{UE}}) \Big)
    \cdot \vec{\mbf{g}}^{\mrm{H}}_{\mrm{BS}}(\psi_{\mrm{BS}}) 
    \mrm{d}\psi_{\mrm{UE}} \mrm{d} \psi_{\mrm{BS}} \nonumber \\
    &= \int_{\psi_{\mrm{BS}}} 
	\mbf{Q}_{\mrm{BS}}^\mrm{T}
    \vec{\mbf{k}}_{\mrm{BS}}(\psi_{\mrm{BS}}) \cdot 
    \vec{P}_{\mrm{h,BS}} (\psi_{\mrm{BS}})  \nonumber \\
    & \cdot \vec{\mbf{k}}^{\mrm{H}}_{\mrm{BS}}(\psi_{\mrm{BS}}) 
	\mbf{Q}_{\mrm{BS}}^\mrm{*}
    \mrm{d} \psi_{\mrm{BS}}.
\end{align}
$\vec{\vec{P}}_{\mrm{h}} (\psi_{\mrm{BS}},\psi_{\mrm{UE}})$ is a joint angular profile, 
which 
is the time-averaged power of the channel response from a certain departure angle 
to a certain arrival angle. 
$\vec{P}_{\mrm{h,BS}} (\psi_{\mrm{BS}}) $ is a marginal angular profile at BS, 
which is determined by the channel response and the beam patterns of UE antennas.
These angular profiles are defined as follows. 
\begin{align}
	& \vec{\vec{P}}_{\mrm{h}} (\psi_{\mrm{BS}},\psi_{\mrm{UE}}) 
	= \mrm{E} \bigg[\left| \vec{\vec{h}}(\psi_{\mrm{BS}},\psi_{\mrm{UE}},t) \right| ^2 \bigg] \\
    & \vec{P}_{\mrm{h,BS}} (\psi_{\mrm{BS}}) \nonumber \\
	&= \int_{\psi_{\mrm{UE}}}
    \vec{\vec{P}}_{\mrm{h}} (\psi_{\mrm{BS}},\psi_{\mrm{UE}}) \cdot 
	\vec{\mbf{g}}^{\mrm{T}}_{\mrm{UE}}(\psi_{\mrm{UE}}) \cdot \vec{\mbf{g}}_{\mrm{UE}}^* (\psi_{\mrm{UE}})
    \mrm{d}\psi_{\mrm{UE}}.
\end{align}
At the UE side, the channel correlation matrix is expressed
 in the same way at the BS.

\subsection{Iterative beam pattern optimization}
\label{sec:Optimization}
\begin{figure}[tb]
    \centering
    \includegraphics[width=0.48\textwidth]{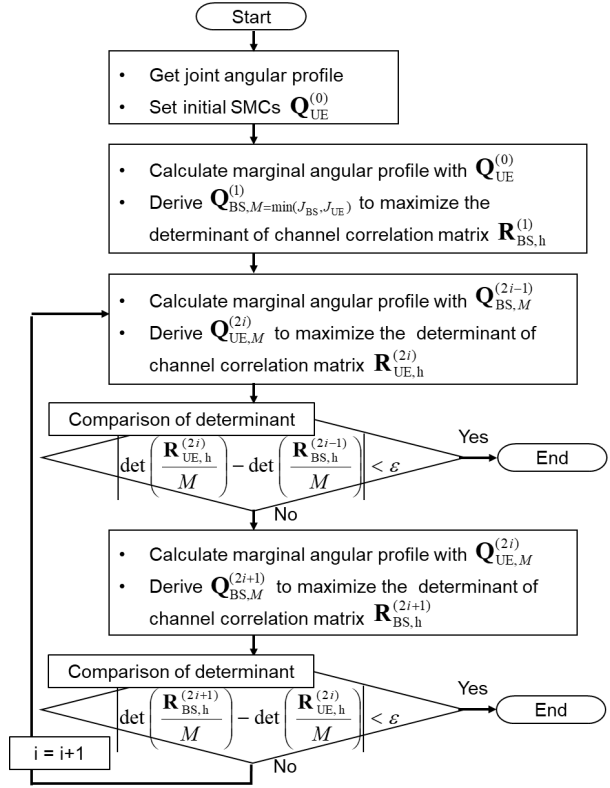}
    \caption{Calculation procedure of iterative optimization at BS and UE.}
    \label{fig:Sequential}
\end{figure}

By using the SU-massive MIMO system model with SME, 
we obtain the optimal beam patterns of BS and UE. 
The optimization method described in \cite{Arai_IEICE}
is expanded in the case of the different antenna volume at BS and UE.
Since the beam patterns are determined by the spherical mode coefficients (SMCs) , 
we introduce the optimization method of SMCs of BS and UE. 
In the method, the analog beamforming weights and beam patterns of antenna elements are considered as 
a matrix of SMCs and far-field pattern functions.
The determinant of the channel correlation matrix can be maximized 
by controlling the matrix of SMCs $\mathbf{Q}_{\mrm{BS},M}$. 
Since the channel correlation matrix is semi-positive definite matrix, 
it can be transformed by the eigenvalue decomposition using the matrix of SMCs.  
The maximum determinant of the channel correlation matrix is expressed 
as follows.
\begin{align}
	& \max \det \mbf{R}_\mrm{BS,h}
	= \max \det (\mbf{Q}^\mrm{T}_{\mrm{BS},M} \mbf{R}_\mrm{BS,sph} \mbf{Q}^{*}_{\mrm{BS},M}) \nonumber \\
	& = \prod_{j=1}^M (\mbf{u}^{\mrm{H}}_{\mrm{BS}j} \mbf{R}_\mrm{BS,sph} \mbf{u}_{\mrm{BS}j})
	= \prod_{j=1}^M \lambda_{\mrm{BS}j}, 
\end{align}
where $\mbf{u}_{\mrm{BS}j}$ is an 
eigenvector and
$\lambda_{\mrm{BS}j} (j=1,\cdots,J)$ is an eigenvalue 
of  the spherical mode correlation matrix 
$\mbf{R}_{\mrm{BS,sph}} \! \in \! \mbb{C}^{J_\mrm{BS} \times J_\mrm{BS}}$. 
The equality is achieved when $\mbf{q}^\mrm{T}_{\mrm{BS}i} \mbf{R}_\mrm{BS,sph} \mbf{q}^*_{\mrm{BS}j}\!=\!0 \ (i \! \neq \! j)$ is satisfied. 
Thus, the vectors to maximize the determinant of the channel correlation matrix are derived 
by the eigenvectors from the first to the $M$-th order of $\mbf{R}_{\mrm{BS,sph}}$
as shown in Fig.\ \ref{fig:Sequential}. 
These calculations should be repeated until the value of objective function converges.
The convergence conditions at BS and UE are indicated respectively as follows.
\begin{align}
	& \left|\det \frac{\mbf{R}_\mrm{a}^{(2i)}}{M} 
	- \det \frac{\mbf{R}_\mrm{b}^{(2i-1)}}{M} \right| < \epsilon \\
	& \left|\det \frac{\mbf{R}_\mrm{b}^{(2i+1)}}{M}
	- \det \frac{\mbf{R}_\mrm{a}^{(2i)}}{M} \right| < \epsilon,
\end{align}
where ``a'' means BS or UE, ``b'' means UE or BS
and  $\epsilon$ is an allowable difference.

\subsection{Rank adaptation}
The rank of the channel correlation matrix depends on the numbers of BS and UE antennas or 
those of BS and UE spherical modes, and the initial condition defined by the angular profile. 
Thus, the optimal rank, i.e., the optimal number of streams, should be derived to maximize the average channel capacity. 
From the results of the analog beamforming weights' selection or iterative calculation, 
the optimal number of streams can be obtained as follows. 
\begin{align}
	M^\mrm{opt} &= \argmax_{M}  \sum_{m=1}^M \bar{C}_m \nonumber \\
	&= \argmax_M  \sum_{m=1}^M \mrm{E} \left[\log_2 \left(1+ \lambda_m \frac{P_\mrm{t}}{M P_\mrm{n}} \right) \right] ,  
\end{align}
where $\lambda_m$ is an eigenvalue of the matrix $\mbf{R}_\mrm{BS,h}$.

\subsection{Convergence of the objective function}
The objective function to maximize the determinant of the channel correlation matrix
converges because it is bounded above and monotonically increasing. 
Since the number of streams is limited, the determinant of the channel correlation matrix is bounded above by the product of a finite number of eigenvalues determined by the number of streams.
Furthermore, when the duality of the channels  is assumed, 
the channel correlation matrix of the $(2n+1)$-th calculation is expressed 
by using $\mbf{Q}_\mrm{BS}^{(2n+1)}$ and $\mbf{R}_\mrm{BS,sph}^{(2n+1)}$
and by using $\mbf{Q}_\mrm{UE}^{(2n)}$ and $\mbf{R}_\mrm{UE,sph}^{(2n+2)}$. 
\begin{eqnarray}
\label{eq:Calc2n+1}
	\mbf{Q}_\mrm{BS}^{(2n+1) \mrm{T}} \mbf{R}_\mrm{BS,sph}^{(2n+1)} \mbf{Q}_\mrm{BS}^{(2n+1) \mrm{*}}
	&=& \mbf{Q}_\mrm{UE}^{(2n) \mrm{T}} \mbf{R}_\mrm{UE,sph}^{(2n+2)}  \mbf{Q}_\mrm{UE}^{(2n) \mrm{*}}
\end{eqnarray}
where superscripts indicate the iteration counts. 
By the $(2n+2)$-th calculation, the matrix of SMCs is derived and the determinant is the same or larger than that of the previous calculation. 
Thus, the determinant is monotonically increasing as follows. 
\begin{eqnarray}
	\det \mbf{Q}_\mrm{UE}^{(2n) \mrm{T}} \mbf{R}_\mrm{UE,sph}^{(2n+2)}  \mbf{Q}_\mrm{UE}^{(2n) \mrm{*}} 
	&\leq&  \prod_{m=1}^{M^{(2n+2)}} \lambda_m^{(2n+2)}  \nonumber \\
	&=& \det \mbf{U}^{(2n+2)\mrm{H}} \mbf{R}_\mrm{UE,sph}^{(2n+2)}  \mbf{U}  \nonumber \\ 
	&=& \det \mbf{Q}_\mrm{UE}^{(2n+2) \mrm{T}} 
	\mbf{R}_\mrm{UE,sph}^{(2n+2)}  \mbf{Q}_\mrm{UE}^{(2n+2) \mrm{*}}, \nonumber \\
\end{eqnarray}
where $\mbf{U}^{(2n+2)}$ and $\lambda_m^{(2n+2)}$ are 
the eigenvectors' matrix and the $m$-th eigenvalue of $\mbf{R}_\mrm{UE,sph}^{(2n+2)}$ respectively.

From the above, the objective function converges 
because it is bounded above and monotonically increasing. 
Additionally, the convergence may be slow when the channels between BS and UE are correlated as shown in Sect. 5.2 of \cite{Arai_IEICE}. In such a case, the calculation is finished based on the iteration counts, elapsed time, and so on.

\subsection{Projection to conditional beam patterns}
The method to derive the semi-optimal beam patterns at BS are introduced
and the same way is used at UE. 
In general, the current distribution is 
derived by solving the following integral equation.
\begin{align}
	\label{eq:q_Integral}
	\mbf{q}_{\mrm{BS},m}^\mrm{opt} \!=\! \int_{V_\mrm{BS}} \!\!
	\vec{J}_{\mrm{BS},m} (r_\mrm{BS}, \theta_\mrm{BS}, \phi_\mrm{BS}) 
	\cdot \vec{\mbf{f}}_\mrm{BS} (r_\mrm{BS}, \theta_\mrm{BS}, \phi_\mrm{BS}) 
	\mrm{d} V_\mrm{BS}, 
\end{align}
where 
$\mbf{q}_{\mrm{BS},m}^\mrm{opt} \! \in \! \mbb{C}^{J_\mrm{BS} \times 1}$ is a vector of SMCs corresponding the $m$-th optimal beam pattern, 
$\vec{J}_{\mrm{BS},m} (r_\mrm{BS}, \theta_\mrm{BS}, \phi_\mrm{BS})$ is the $m$-th current distribution on the antenna surface with volume $V_\mrm{BS}$
and $\vec{\mbf{f}}_\mrm{BS} (r, \theta, \phi) \! \in \! \mbb{C}^{J_\mrm{BS} \times 1}$ is a vector of spherical wave functions 
representing radial standing waves at a location $(r, \theta, \phi)$. 
By using Galerkin method, which is one of 
the methods for solving the integral equation, 
Eq.\ (\ref{eq:q_Integral}) is represented as a linear equation as shown in \cite{Arai_IEICE}, given by
\begin{align}
	\label{eq:qZa}
	\mbf{q}_{\mrm{BS},m}^\mrm{opt} = \mbf{Z}_\mrm{BS} \mbf{a}_{\mrm{BS},m}, 
\end{align}
where $\mbf{Z}_\mrm{BS} \! \in \! \mbb{C}^{J_\mrm{BS} \times L}$ is a transformation matrix 
from the space of antenna surface to the far field 
and $\mbf{a}_{\mrm{BS},m} \! \in \! \mbb{C}^{L \times 1}$ is a vector of the $m$-th current distribution coefficients. 
By using a Moore-Penrose inverse matrix $\mbf{Z}_\mrm{BS}^+ \! \in \! \mbb{C}^{L \times J_\mrm{BS}}$, 
the current distribution coefficients are derived as 
the least squares and minimum norm solution. 
\begin{align}
	\label{eq:aZ+q}
	\mbf{a}_{\mrm{BS},m} = \mbf{Z}_\mrm{BS}^+ \mbf{q}_{\mrm{BS},m}^\mrm{opt}. 
\end{align}
The SMCs' vector of the semi-optimal beam pattern radiated from the antenna surface 
is expressed as follows. 
\begin{align}
	\mbf{q}_{\mrm{BS},m}^\mrm{semi} = \mbf{Z}_\mrm{BS} \mbf{a}_{\mrm{BS},m}
	= 
	\mbf{Z}_\mrm{BS} \mbf{Z}^+ \mbf{q}_{\mrm{BS},m}^\mrm{opt}. 
\end{align}
$\mbf{Z}_\mrm{BS} \mbf{Z}_\mrm{BS}^+$ is an orthogonal projection operator \cite{Baksalary}, 
thus the SMCs' vector of the optimal beam pattern is projected to that of the conditional beam pattern
and 
the semi-optimal beam pattern is obtained as corresponding to the current distribution of the least squares and the minimum norm solution 
among the conditional beam patterns that can be radiated from the antenna surface.

\begin{figure}[tb]
    \centering
    \includegraphics[width=0.45\textwidth]{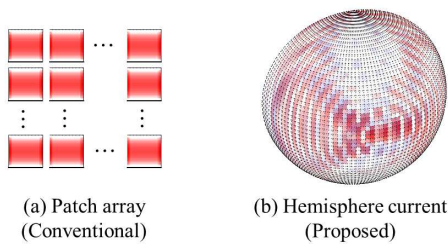}
    \caption{Examples of current distributions of conventional array and proposed current surface.}
    \label{fig:CurrentExample}
\end{figure}

Table\ \ref{tbl:CurrentSurface} shows examples of 
the assumed antenna surface for the BS antenna
defined by a radius of the spherical surface $R$
and a range of angle $\theta_\mrm{c}, \phi_\mrm{c}$.
Three cases of surfaces are considered such as Plane, 1/32-sphere and Hemisphere. 
They are included in the sphere with the radius $r_{0,\mrm{BS}}$,  
which is the same volume as the 2-dimensional square array antennas in 
the cases of full-array and sub-array.
Examples of configurations and current distributions 
are shown in Fig.\ \ref{fig:CurrentExample} 
by using the conventional planar patch array 
and the hemisphere antenna surface that we proposed. 

\begin{table*}[tb]
	\begin{center}
	\caption{Examples of antenna surfaces.}
	\label{tbl:CurrentSurface}
		\begin{tabular}{|c||c|c|c|}
		\hline
		& 
		Plane & 1/32-sphere & Hemisphere  \\
		\hline
		\hline
		& & & \\
		& \begin{minipage}{0.25\textwidth}
      		\centering
      		\includegraphics[width=0.45\textwidth]{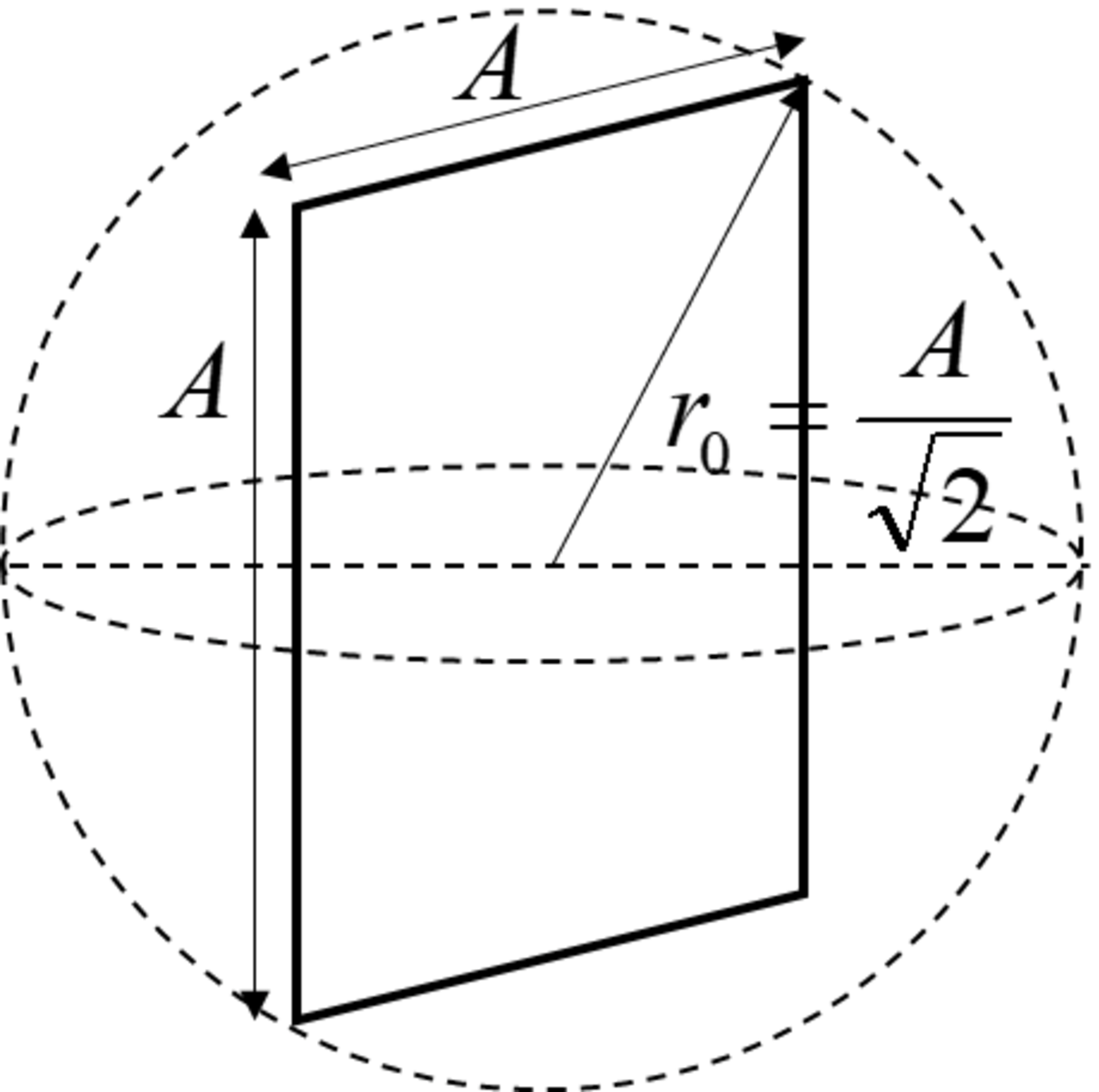}
    	\end{minipage}
		& 
		\begin{minipage}{0.3\textwidth}
      		\centering
      		\includegraphics[width=0.85\textwidth]{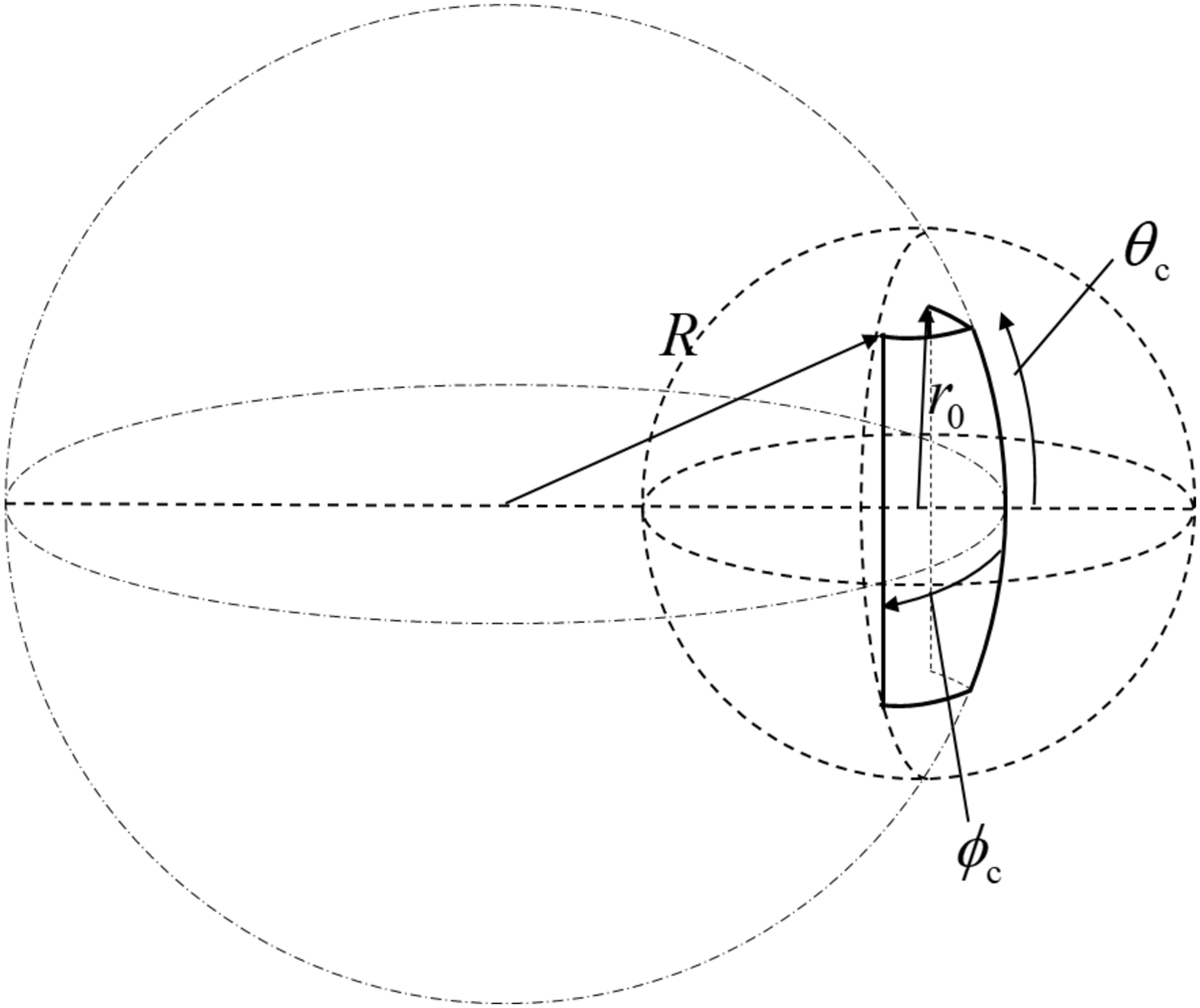}
    	\end{minipage}
		&
		\begin{minipage}{0.25\textwidth}
      		\centering
      		\includegraphics[width=0.45\textwidth]{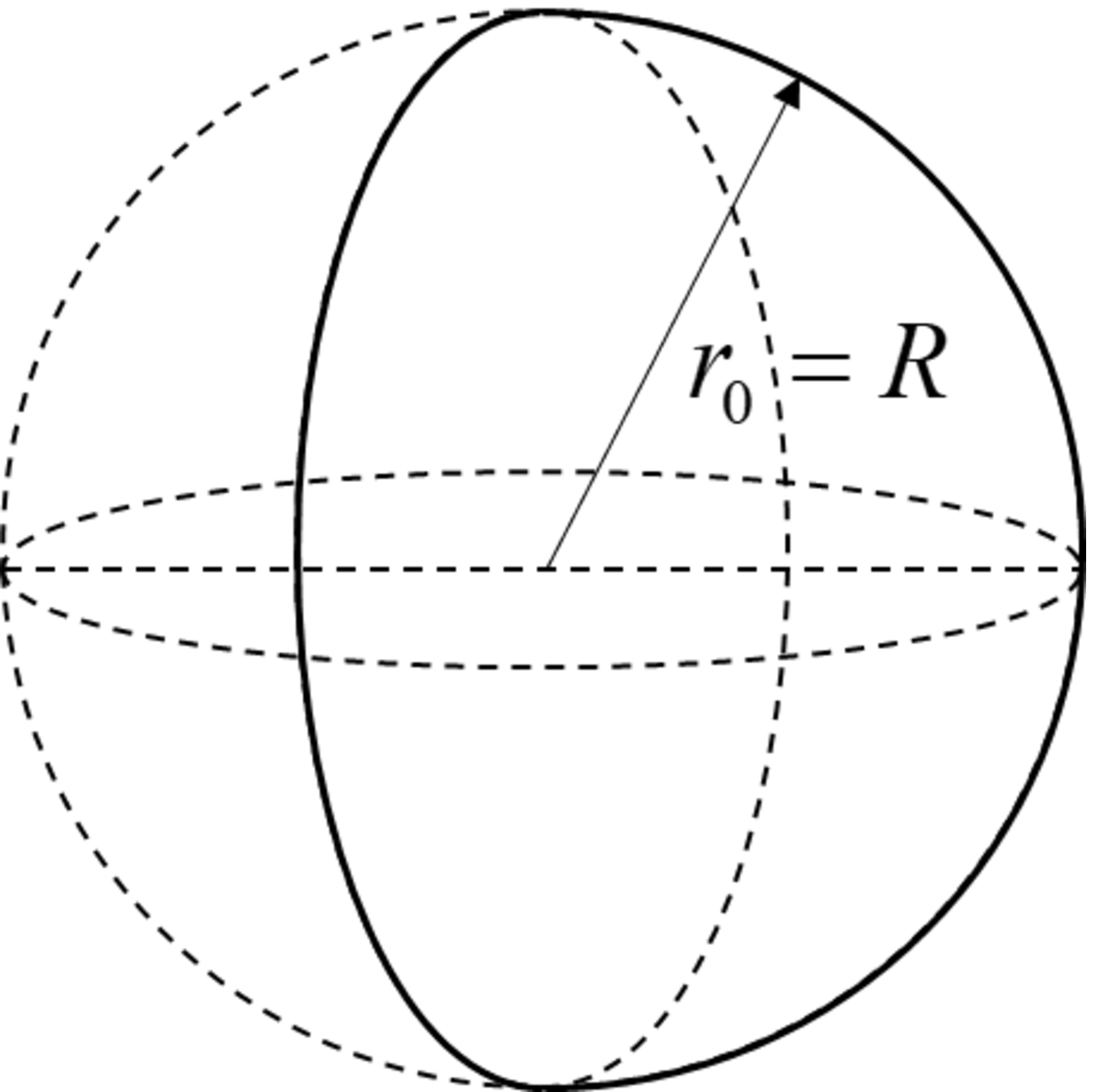}
    	\end{minipage}
		\\ 
		& & & \\
		\hline
		$R$ & -- & $\max \{ r_0 / (\sqrt{2} \sin\theta_c), r_0 / ( \sqrt{2} \sin\phi_c) \}$ & $r_0$ \\
		\hline
		($\theta_\mrm{c}$, $\phi_\mrm{c}$) 
		& --
		& ($\pi / 8$, $\pi / 8$)
		& ($\pi / 2$, $\pi / 2$) \\
		\hline
		\end{tabular}
	\end{center}
\end{table*}

\section{Numerical analysis}
In this section, the beam patterns by using OBPB
and conventional hybrid beamforming are derived and discussed.

\subsection{Analysis condition}
\begin{figure}[tb]
    \centering
    \includegraphics[width=0.36\textwidth]{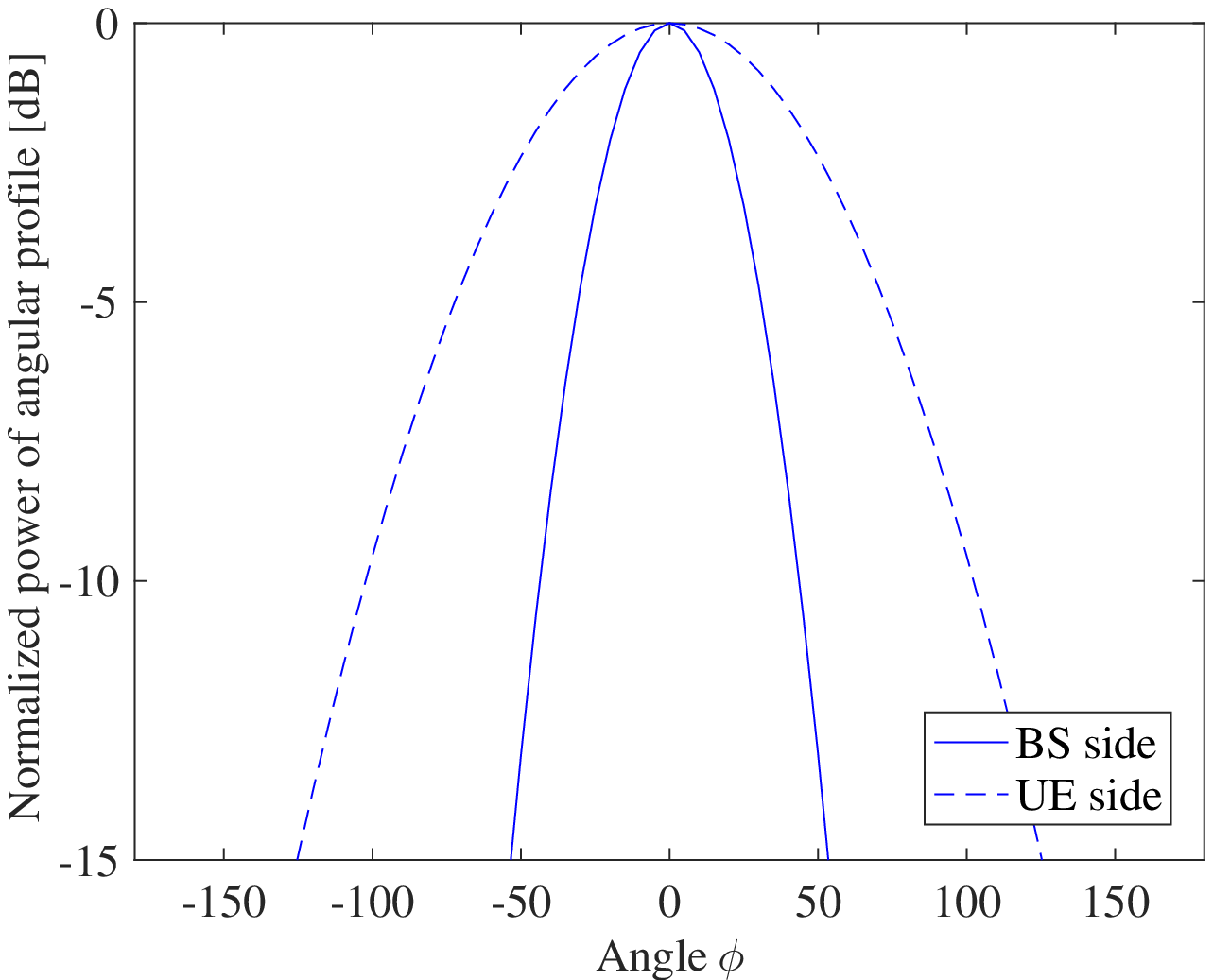}
    \caption{Angular profile in $\phi$-plane.}
    \label{fig:AngularProfile_phi}
    \centering
    \includegraphics[width=0.36\textwidth]{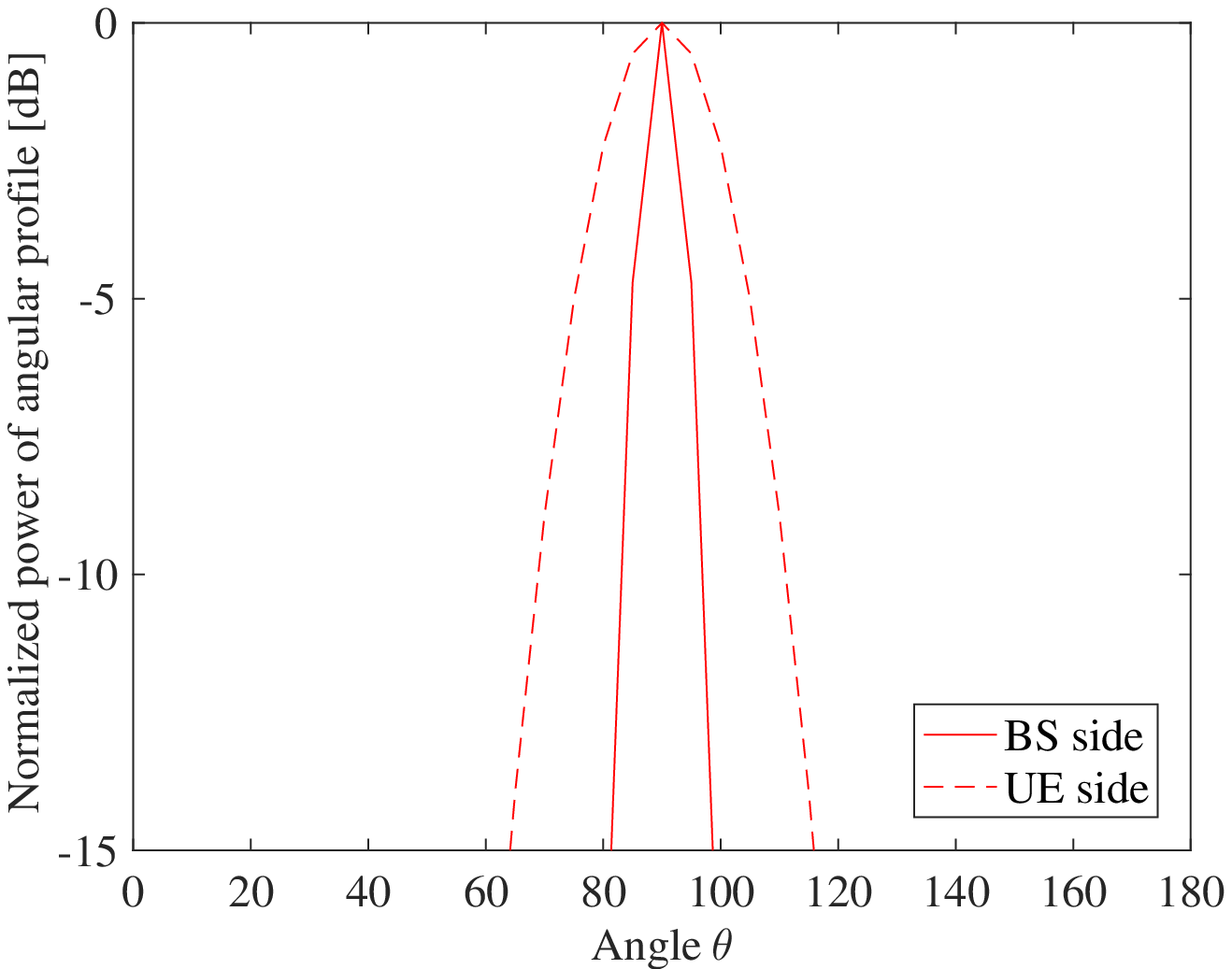}
    \caption{Angular profile in $\theta$-plane.}
    \label{fig:AngularProfile_theta}
\end{figure}

Conditions of analysis are shown in Table\ \ref{tbl:AnalysisCondition}
and the angular profile is defined by using a multivariate Gaussian distribution
as shown in Figs.\ \ref{fig:AngularProfile_phi}, \ref{fig:AngularProfile_theta} 
and Table\ \ref{tbl:AngularProfiles} with only $\theta$ polarization component. 
In the analysis, the $\phi$-plane and $\theta$-plane indicate the $xy$-plane and $xz$-plane as shown in Fig.\ \ref{fig:SphericalCoordinate} respectively.
$\lambda_0$ is a wavelength,  
$\mbf{P}$ is a correlation matrix between each variates, 
the standard deviations of $\theta$ components are $\sigma_{\mrm{BS}, \theta}, \sigma_{\mrm{UE}, \theta}$, 
and the standard deviations of $\phi$ components are $\sigma_{\mrm{BS}, \phi}, \sigma_{\mrm{UE}, \phi}$.
The mean and correlation matrix between 
variables are defined 
as 3GPP UMa NLOS model at 30 GHz \cite{3GPP}.

In the conventional hybrid beamforming case, 
the 2-dimensional square array antennas are used
and each element has a beam pattern defined in \cite{3GPP}.
To compare the characteristics of the proposed and conventional hybrid beamforming 
in the same antenna volume, 
the antenna area of the patch array at the UE side is a square having sides with a fixed length $A_\mrm{UE}$ 
and the elements are allowed to be overlapped when the number of UE antennas becomes large. 
In the case of sub-array configuration, the number of antenna elements in the sub-array 
is chosen from 10 types of sub-array configurations to maximize the average channel capacity, 
such as ($N_\mrm{BS,sub,V} \times N_\mrm{BS,sub,H}$)
= ($1 \times 4$), ($2 \times 2$), ($4 \times 1$), 
($1 \times 8$), ($2 \times 4$), ($4 \times 2$), ($8 \times 1$), 
($2 \times 8$), ($4 \times 4$), ($8 \times 2$). 

\begin{figure}[tb]
    \centering
    \includegraphics[width=0.45\textwidth]{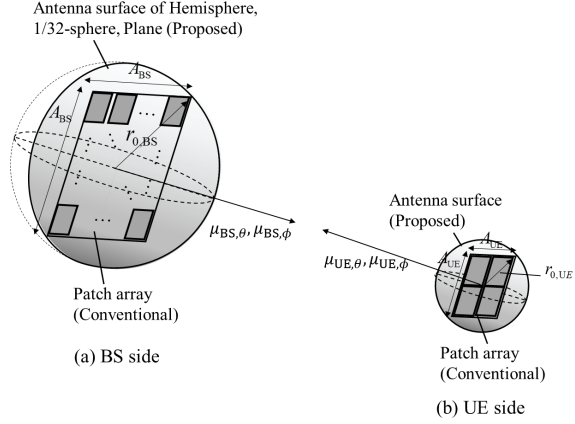}
    \caption{Antenna configurations for proposed and conventional methods.}
    \label{fig:AntennasForAnalysis}
\end{figure}

The current surfaces of the proposed cases
are shown in Table\ \ref{tbl:CurrentSurface}. 
They are defined as the same antenna volume as the conventional array configurations, 
which are defined by the same radii of BS and UE antenna volume $r_\mrm{0,BS}, r_\mrm{0,UE}$. 
The relationship between 
the proposed and the conventional antenna configurations is 
shown in Fig.\ \ref{fig:AntennasForAnalysis}, 
where they face each other with a direction of 
$(\mu_{\mrm{BS}, \theta}, \mu_{\mrm{BS}, \phi})=(90\ \mrm{[deg.]}, 0\ \mrm{[deg.]})$ at BS
and that of 
$(\mu_{\mrm{UE}, \theta}, \mu_{\mrm{UE}, \phi})=(90\ \mrm{[deg.]}, 0\ \mrm{[deg.]})$ 
at UE.

\begin{table}[tb]
	\begin{center}
	\caption{Analysis condition.}
	\label{tbl:AnalysisCondition}
		\begin{tabular}{|c||c|}
		\hline
		& $N_\mrm{BS}= 64$ \\
		Number of antennas & $N_\mrm{UE}= 4, 9, 16, 25, 36, 49$ \\
		& $N_\mrm{BS,sub}= 4, 8, 16$ \\
		\hline
		Spacing of BS antenna elements & $\frac{\lambda_0}{2}$ \\
		\hline
		Spacing of UE antenna elements & $\frac{1}{\sqrt{N_\mrm{UE}}-1} \cdot \frac{\lambda_0}{2}$ \\
		\hline
		Antenna size & $\frac{\lambda_0}{2}$ \\
		\hline
		Beam interval coefficient & $a$ = 4 \\
		\hline
		Radius of antenna volume & $r_\mrm{0,BS}
		= \left(\frac{ 7 \lambda_0}{2} + \frac{\lambda_0}{2} \right) / \sqrt{2} $ \\
		  & $r_\mrm{0,UE} 
		= \left(\frac{ \lambda_0}{2} + \frac{\lambda_0}{2} \right) / \sqrt{2} $ \\
		\hline	
		Number of spherical modes & $J_\mrm{BS}=646$ \\
		 & $J_\mrm{UE}=48$ \\
		\hline	
		Initial beam pattern & omni-directional pattern \\
		for iterative optimization & \\
		\hline
		Initial number of streams & 1 \\
		for iterative optimization & \\
		\hline
		Received SNR with & -12 dB \\
		omni directivities in SISO &  \\
		\hline
		Basis function & Dirac delta function \\
		\hline
		& 
		1\% of the difference \\
		Allowable difference &
		between the last value and \\ 
		&
		the previous value \\
		\hline
		\end{tabular}
	\end{center}
\end{table}

\begin{table}[tb]
	\begin{center}
	\caption{Parameters of angular profile.}
	\label{tbl:AngularProfiles}
		\begin{tabular}{|c||c|}
		\hline
		& $\mu_{\mrm{BS}, \theta} =$ 90 [deg.] \\
		Angle of departure & $\mu_{\mrm{BS}, \phi} =$ 0 [deg.] \\
		and arrival & $\mu_{\mrm{UE}, \theta} =$ 90 [deg.] \\ 
		& $\mu_{\mrm{UE}, \phi} =$ 0 [deg.] \\
		\hline
		& $\sigma_{\mrm{BS}, \theta} =$ 4 [deg.] \\
		Angular spread of & $\sigma_{\mrm{BS}, \phi} =$ 21 [deg.] \\
		departure and arrival & $\sigma_{\mrm{UE}, \theta} =$ 11 [deg.] \\
		& $\sigma_{\mrm{UE}, \phi} =$ 48 [deg.] \\
		\hline
		Covariance matrix $\mbf{P}$ 
		&  $\left[ 
			\begin{array}{cccc}
			1 & 0.3 & 0 & 0.2 \\
			0.3 & 1 & 0.1 & 0.4 \\
			0 & 0.1 & 1 & 0 \\
			0.2 & 0.4 & 0 & 1\\
			\end{array} 
			\right]$ 
		\\ 
		\hline
		Polarization & Only $\theta$ polarization \\
		\hline
		\end{tabular}
	\end{center}
\end{table}

\subsection{Beam patterns}
\begin{figure}[htb]
    \centering
    \includegraphics[width=0.36\textwidth]{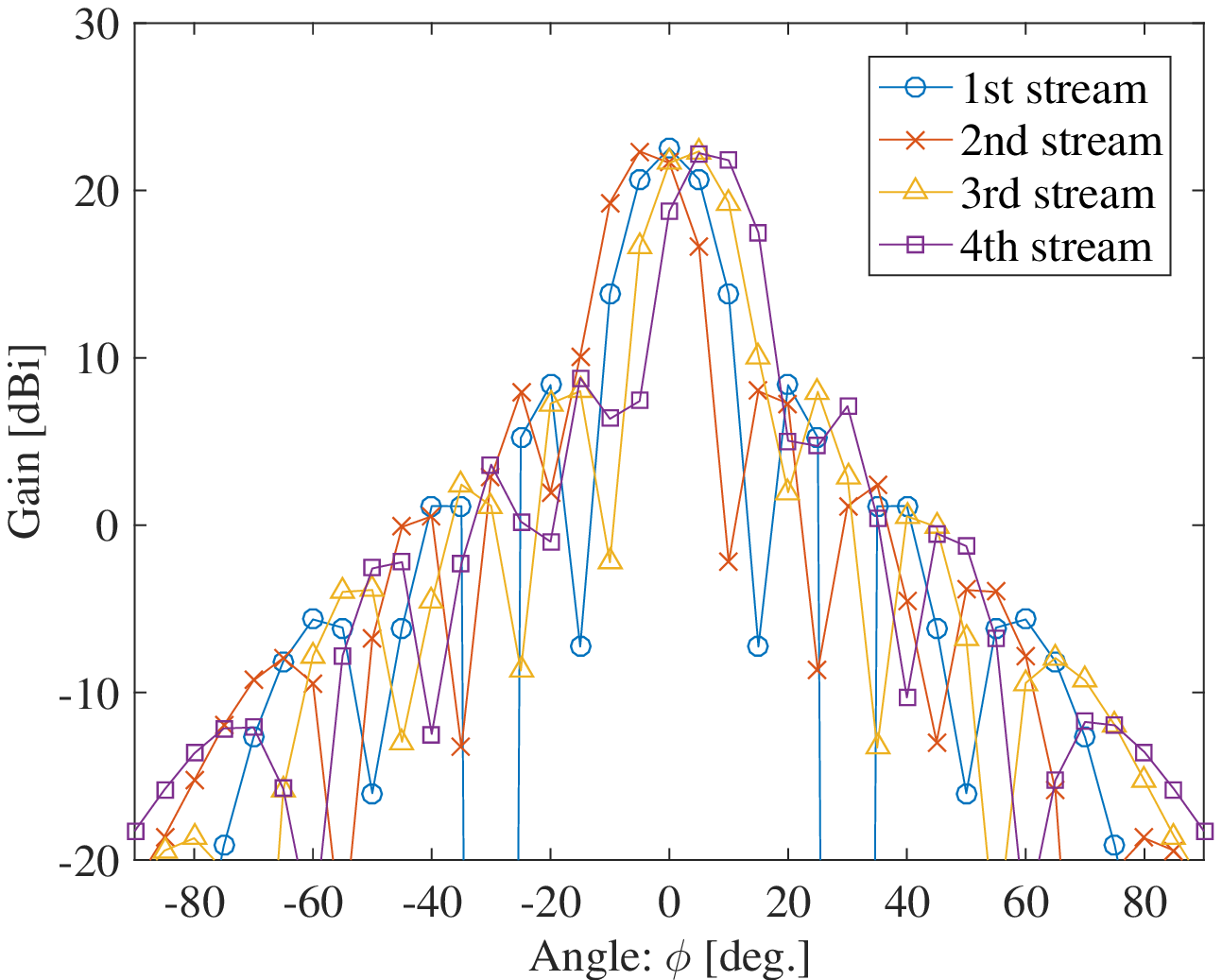}
    \caption{Beam patterns of Full-array in $\phi$-plane derived by Received power ($N_\mrm{UE}=4$).}
    \label{fig:Full1_D_phi}
    \centering
    \includegraphics[width=0.36\textwidth]{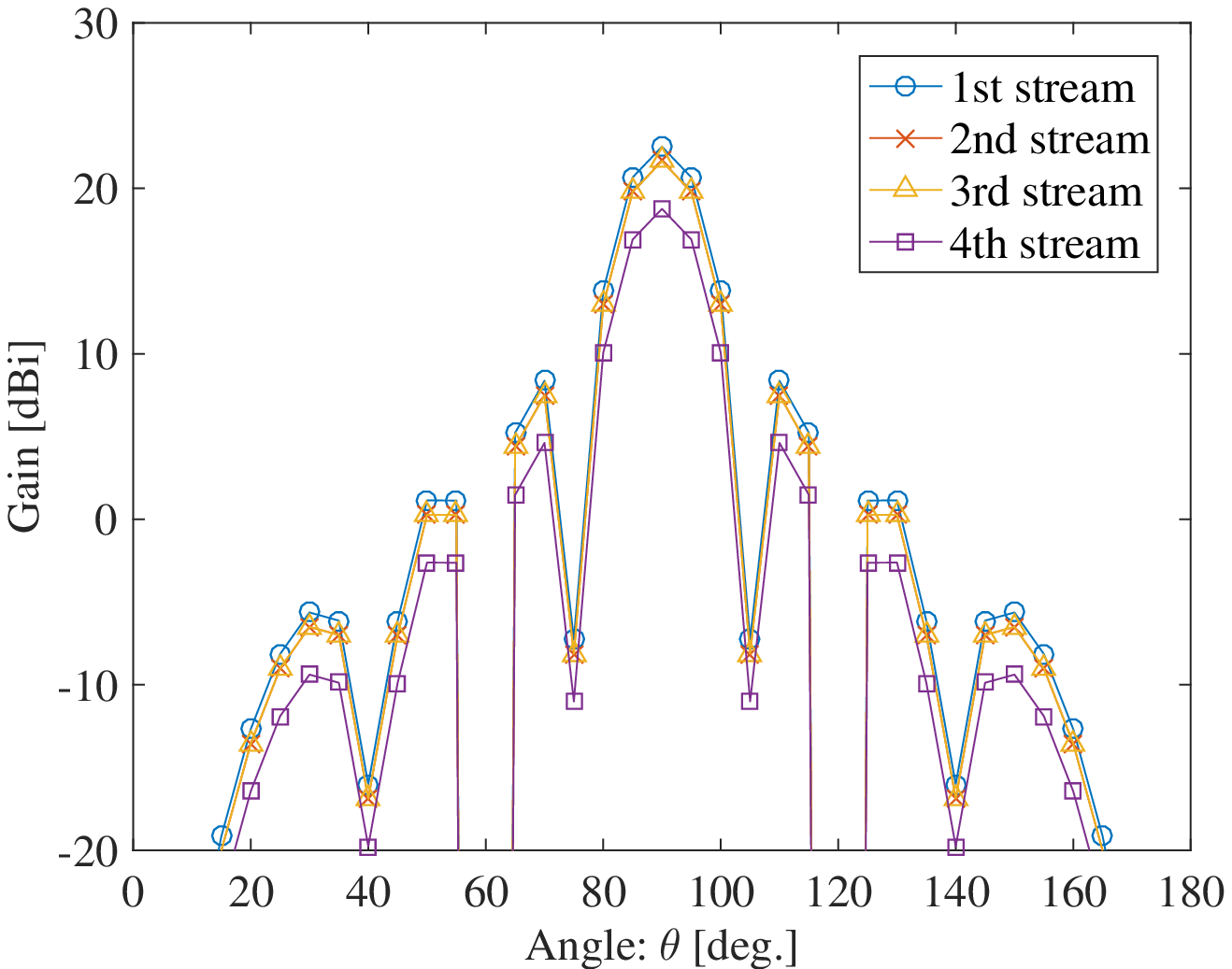}
    \caption{Beam patterns of Full-array in $\theta$-plane derived by Received power ($N_\mrm{UE}=4$).}
    \label{fig:Full1_D_theta}
    \centering
    \includegraphics[width=0.36\textwidth]{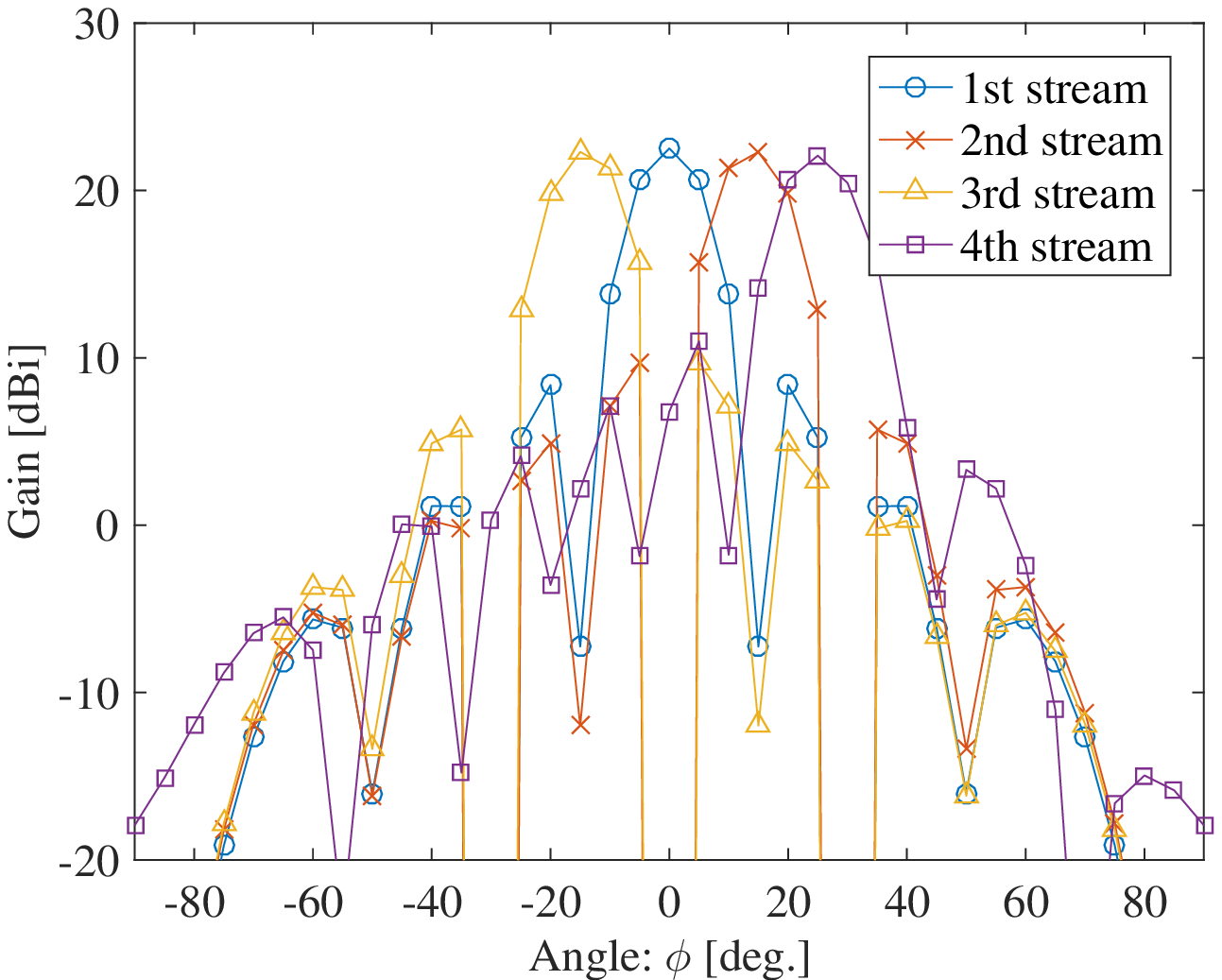}
    \caption{Beam patterns of Full-array in $\phi$-plane derived by Determinant ($N_\mrm{UE}=4$).}
    \label{fig:Full2_D_phi}
    \centering
    \includegraphics[width=0.36\textwidth]{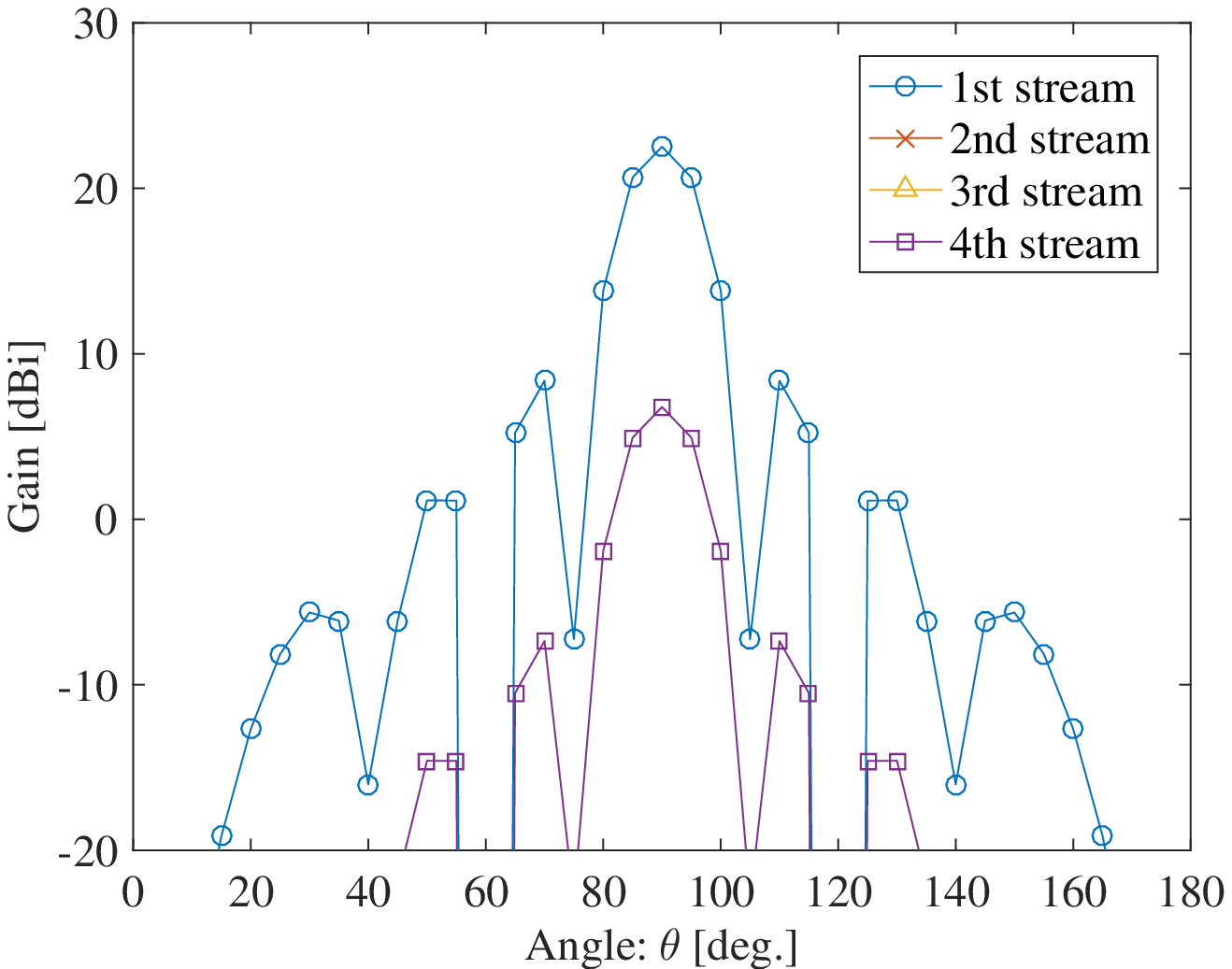}
    \caption{Beam patterns of Full-array in $\theta$-plane derived by Determinant ($N_\mrm{UE}=4$).}
    \label{fig:Full2_D_theta}
\end{figure}

The beam patterns of the full-array configuration (Full-array) are derived as shown in Figs.\ \ref{fig:Full1_D_phi}, \ref{fig:Full1_D_theta}, \ref{fig:Full2_D_phi} and \ref{fig:Full2_D_theta}
in the case of $N_\mrm{UE}=4$.
By using the received power of each stream for the weight selection, the patterns near 
the peak of the angular profile are chosen. 
It causes the degradation of the average channel capacity
due to the high correlation between the beam patterns. 
On the other hand, by using the determinant of the channel correlation matrix,
the average channel capacity does not degrade compared with the previous case
because sufficiently separated and low correlated patterns are selected 
by considering the determinant. 
However,
as the number of streams increases, the peaks of the selected patterns are away from the peaks of the given angular profile.
Thus, it causes the loss of the received power and the capacity degradation. 

\begin{figure}[htb]
    \centering
    \includegraphics[width=0.36\textwidth]{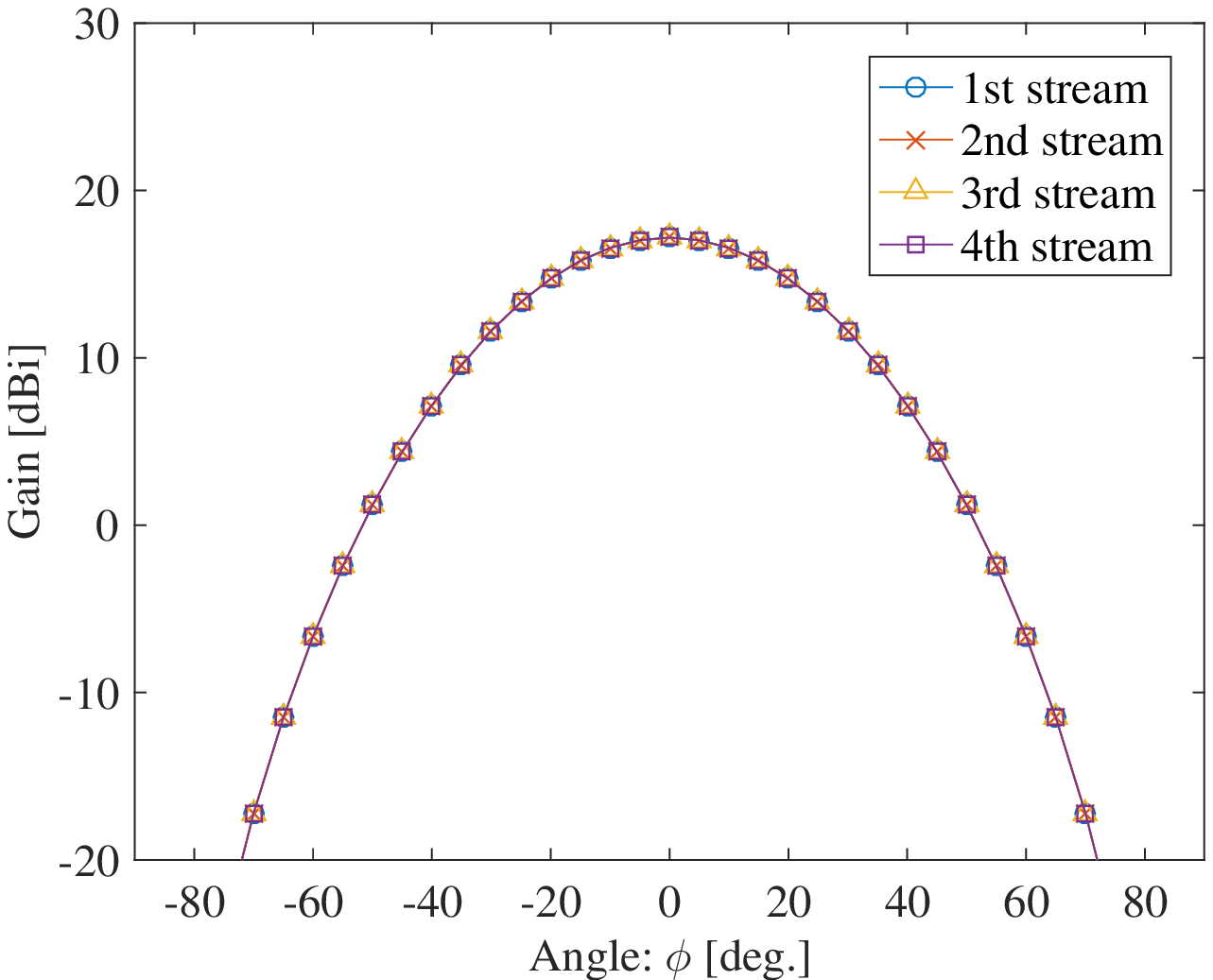}
    \caption{Beam patterns of Sub-array in $\phi$-plane ($N_\mrm{UE}=4$).}
    \label{fig:Sub_D_phi}
    \centering
    \includegraphics[width=0.36\textwidth]{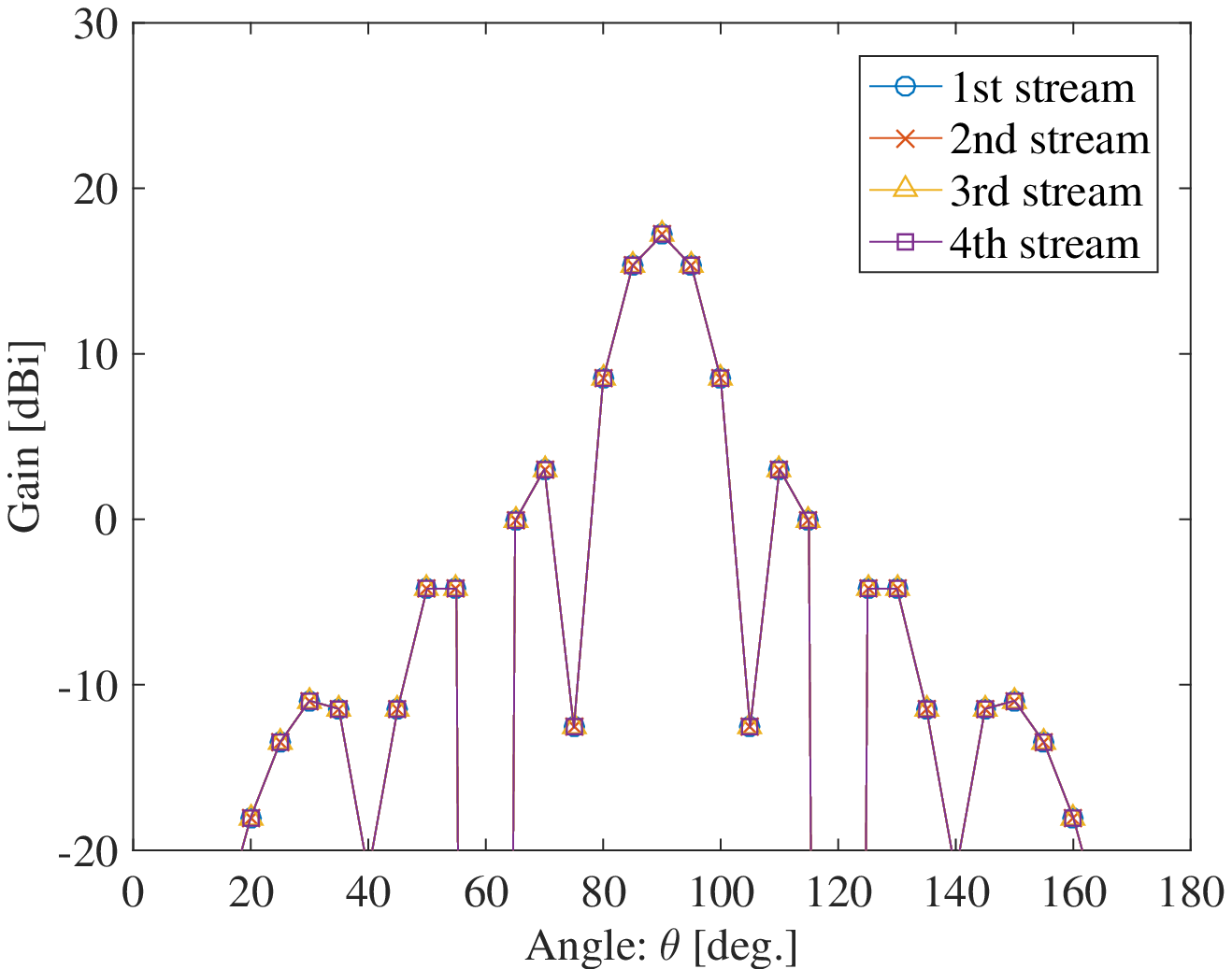}
    \caption{Beam patterns of Sub-array in $\theta$-plane ($N_\mrm{UE}=4$).}
    \label{fig:Sub_D_theta}
    \centering
    \includegraphics[width=0.36\textwidth]{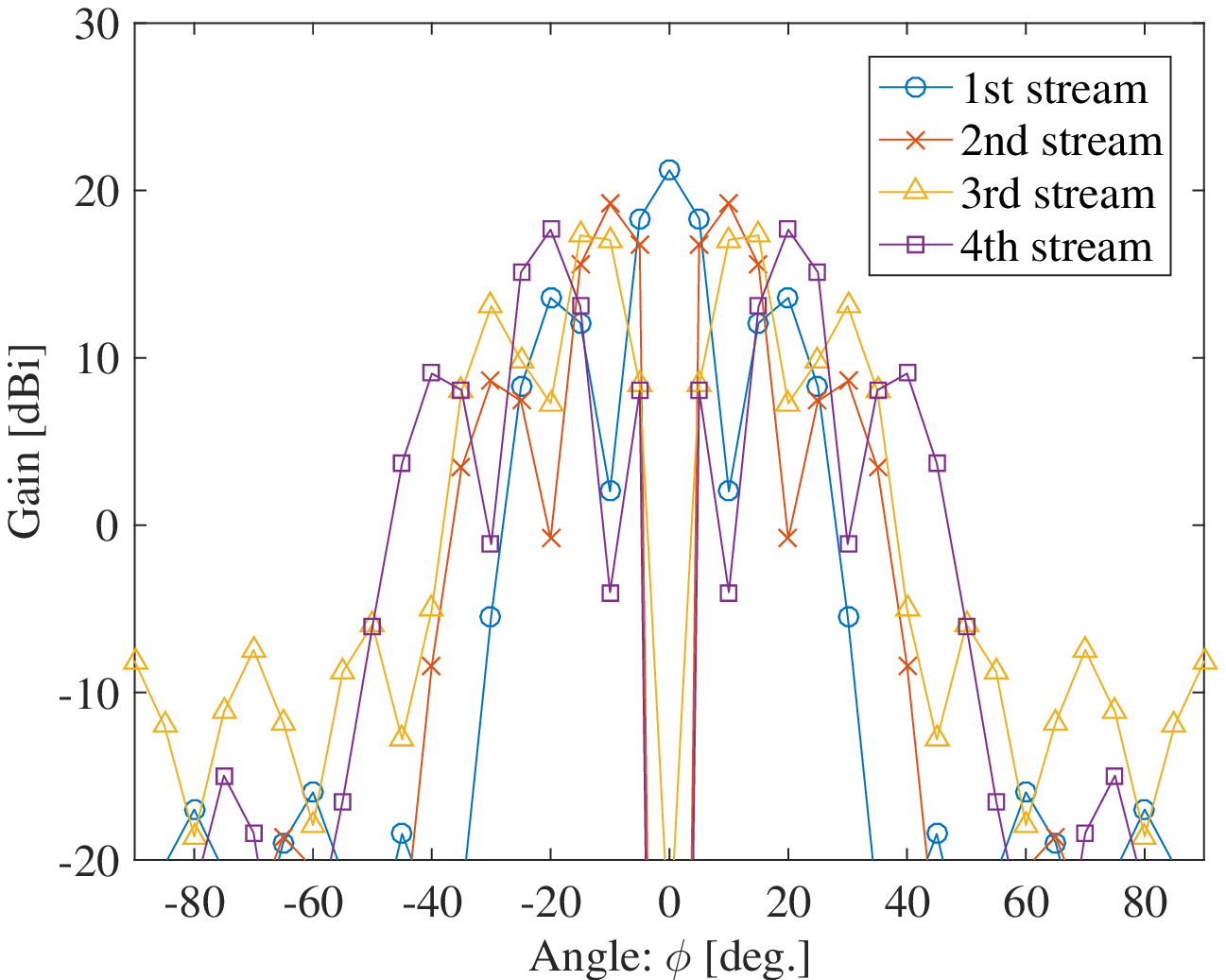}
    \caption{Semi-optimal beam patterns of Plane in $\phi$-plane.}
    \label{fig:Plane_D_phi}
    \centering
    \includegraphics[width=0.36\textwidth]{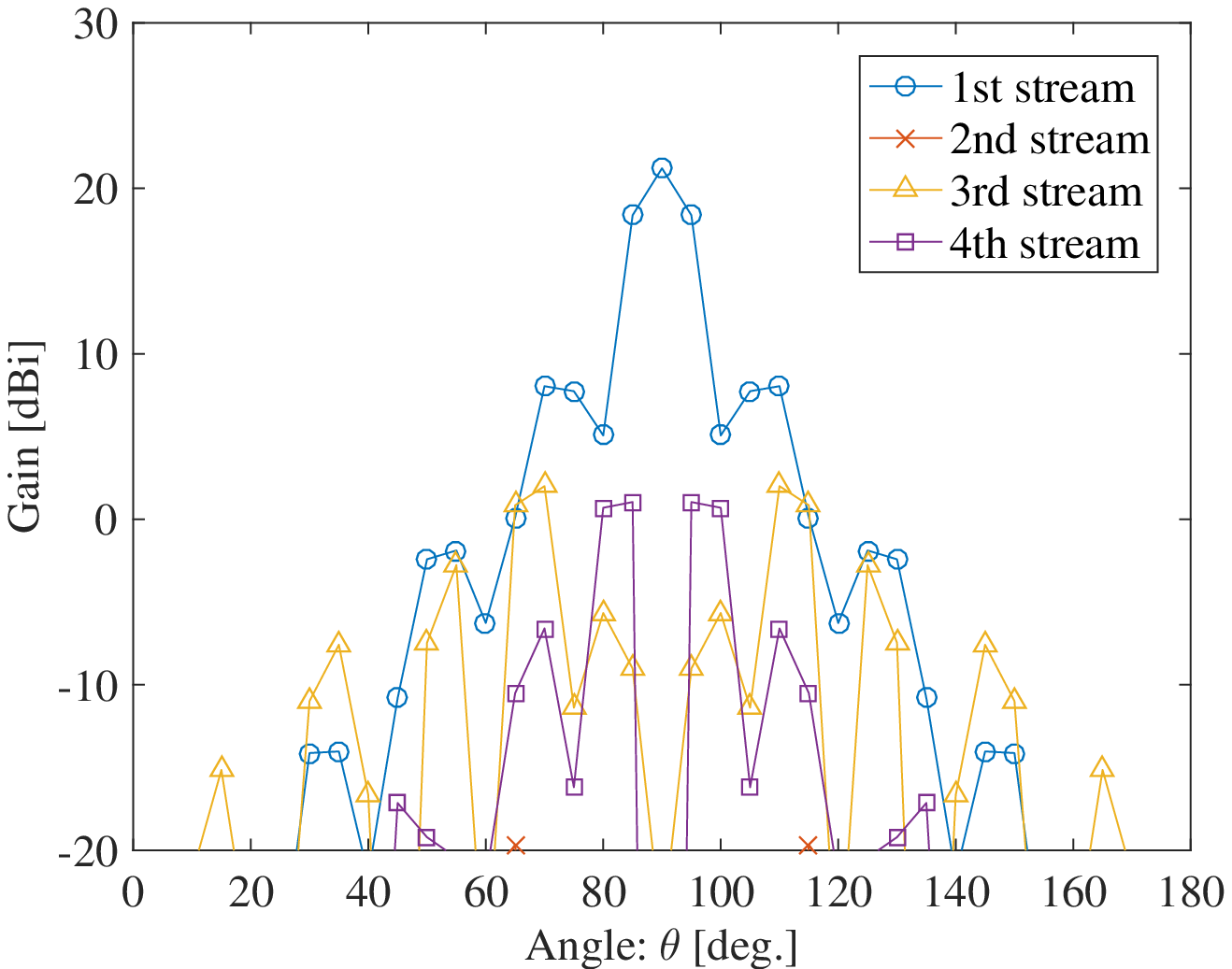}
    \caption{Semi-optimal beam patterns of Plane in $\theta$-plane.}
    \label{fig:Plane_D_theta}
\end{figure}

Next, the optimal beam patterns of the sub-array configuration (Sub-array) is shown in Figs.\ \ref{fig:Sub_D_phi} and \ref{fig:Sub_D_theta} in the case of $N_\mrm{UE}=4$.
The same beam patterns are selected in all groups of sub-arrays by using either the received power and the determinant.  
The number of antenna elements of the sub-array group is derived as 
($N_\mrm{BS,sub,V} \times N_\mrm{BS,sub,H}$)= ($8 \times 2$) in the case of $N_\mrm{UE}=4$, 
($N_\mrm{BS,sub,V} \times N_\mrm{BS,sub,H}$)= ($4 \times 2$) in the case of $N_\mrm{UE}=9$, 
and ($N_\mrm{BS,sub,V} \times N_\mrm{BS,sub,H}$)= ($4 \times 1$) in the other cases. 
The angular profile in $\theta$-plane is narrower than that of $\phi$-plane in the analysis. 
Thus, the beam patterns in $\theta$-plane should be narrow as well
and the sub-array is better to be a vertical array than a horizontal array. 
Furthermore, it is found that the same beam patterns are chosen in this case. 
From the results, the beam patterns are sufficiently decorrelated 
even though using the same analog beamforming weight vector
because the effective 
radio wave sources are  
separated from with each other.

\begin{figure}[htb]
    \centering
    \includegraphics[width=0.36\textwidth]{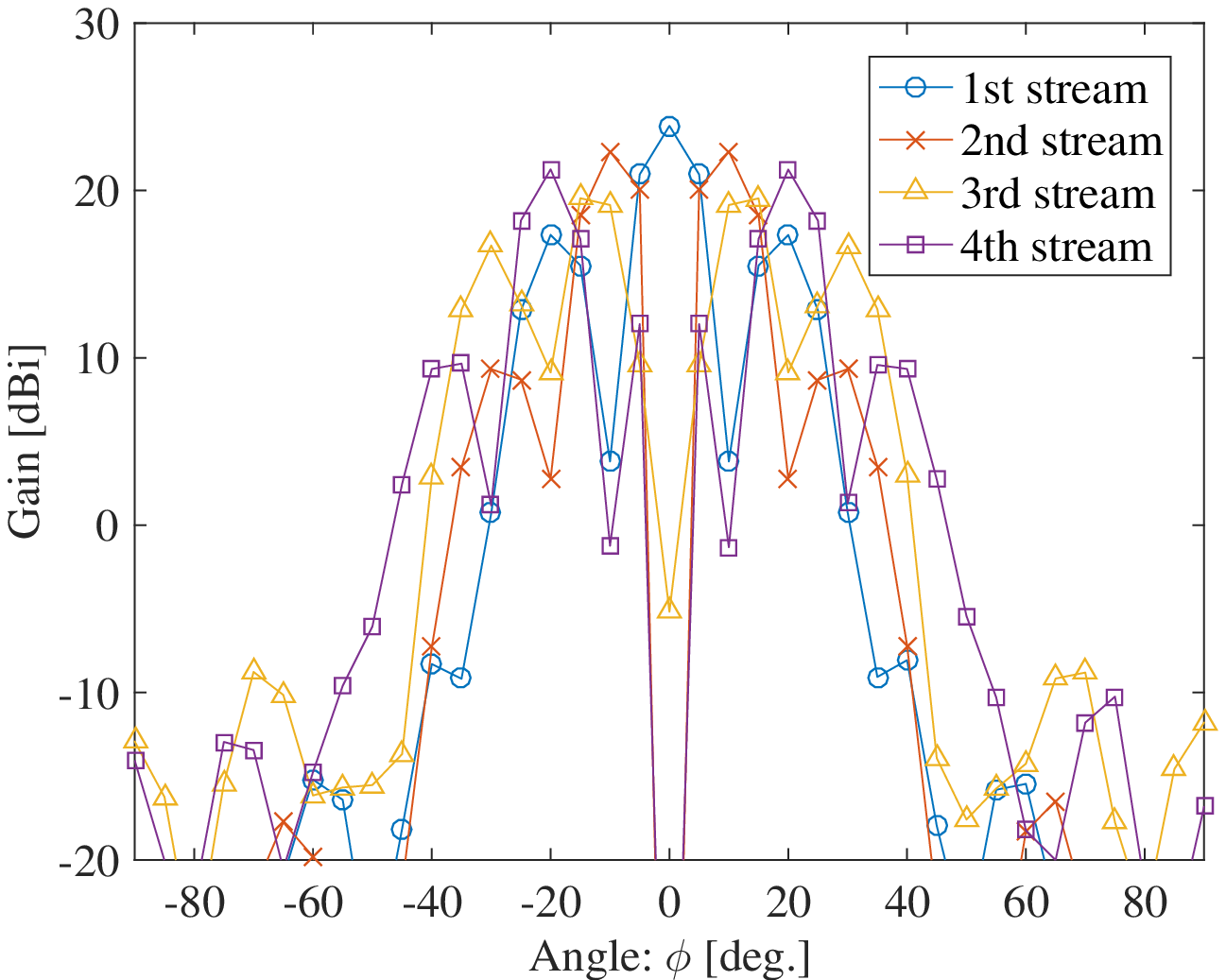}
    \caption{Semi-optimal beam patterns of 1/32-sphere in $\phi$-plane.}
    \label{fig:Sphere_D_phi}
    \centering
    \includegraphics[width=0.36\textwidth]{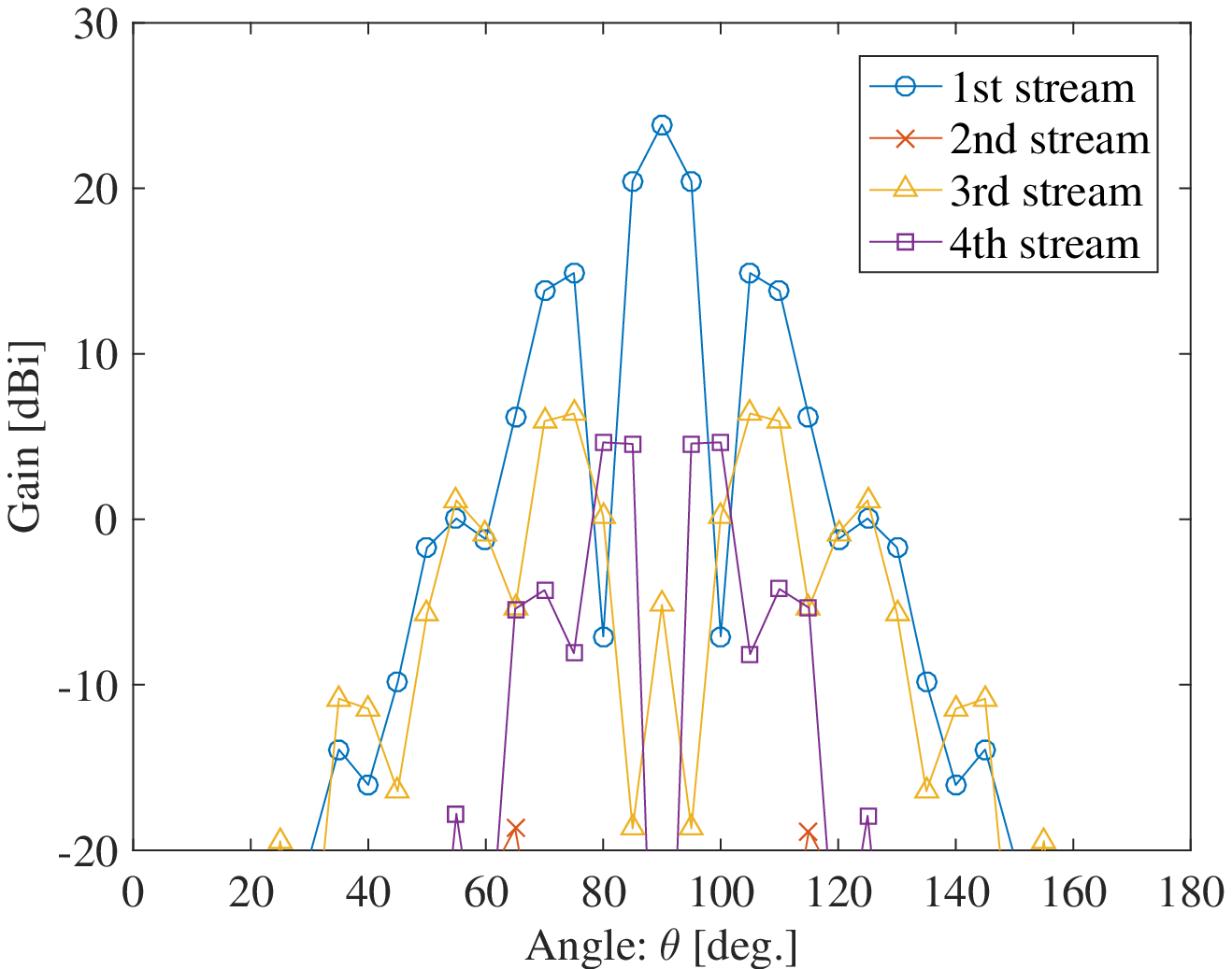}
    \caption{Semi-optimal beam patterns of 1/32-sphere in $\theta$-plane.}
    \label{fig:Sphere_D_theta}
    \centering
    \includegraphics[width=0.36\textwidth]{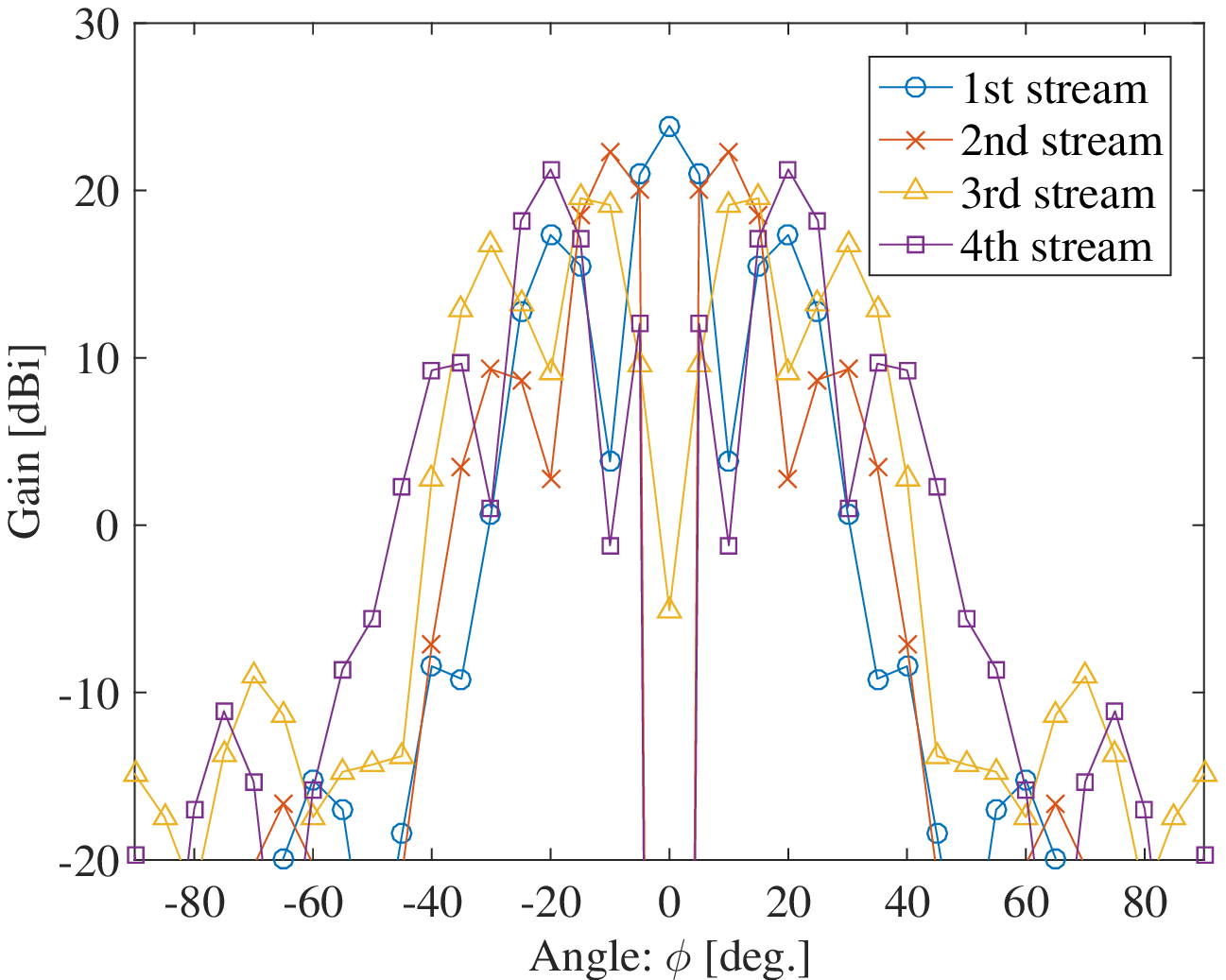}
    \caption{Semi-optimal beam patterns of Hemisphere in $\phi$-plane.}
    \label{fig:SME_D_phi}
    \centering
    \includegraphics[width=0.36\textwidth]{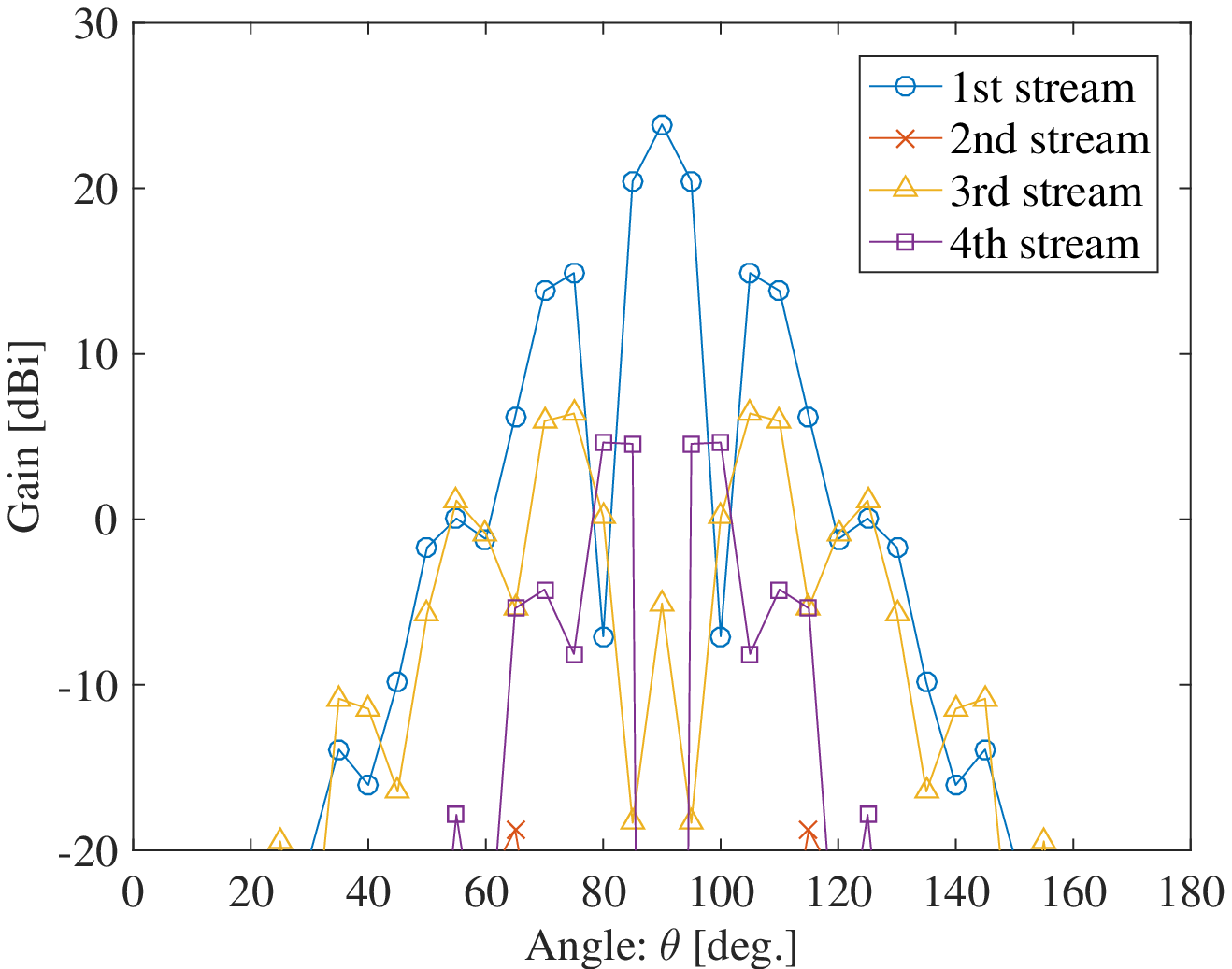}
    \caption{Semi-optimal beam patterns of Hemisphere in $\theta$-plane.}
    \label{fig:SME_D_theta}	
\end{figure}

By using OBPB on  the spherical surface, 
the optimal beam patterns are derived as shown 
in Fig.\ \ref{fig:Plane_D_phi} to Fig.\ \ref{fig:SME_D_theta}.
It is found that the pattern of the 1st stream has its peak
and the other patterns have null towards the peak of the angular profile. 
In the case of Plane, the gains degrade 
because 
the planar antenna surface cannot make some beam patterns
due to constraints of the conditional current distributions on the surface. 
When the current surface becomes curved from Plane to 1/32-sphere and Hemisphere,
the power of the beam pattern is directed to the peak of the angular profile. 
Thus, the loss of the transmission power becomes small while having low correlation
and the average channel capacity can be improved.
As the current surface is curved as 1/32-sphere and Hemisphere, 
the complexity of the derived beam patterns becomes high
and they have low 
gains of side lobes compared to Plane  
because the various directions of currents are achieved. 


\begin{figure}[htb]
    \centering
    \includegraphics[width=0.45\textwidth]{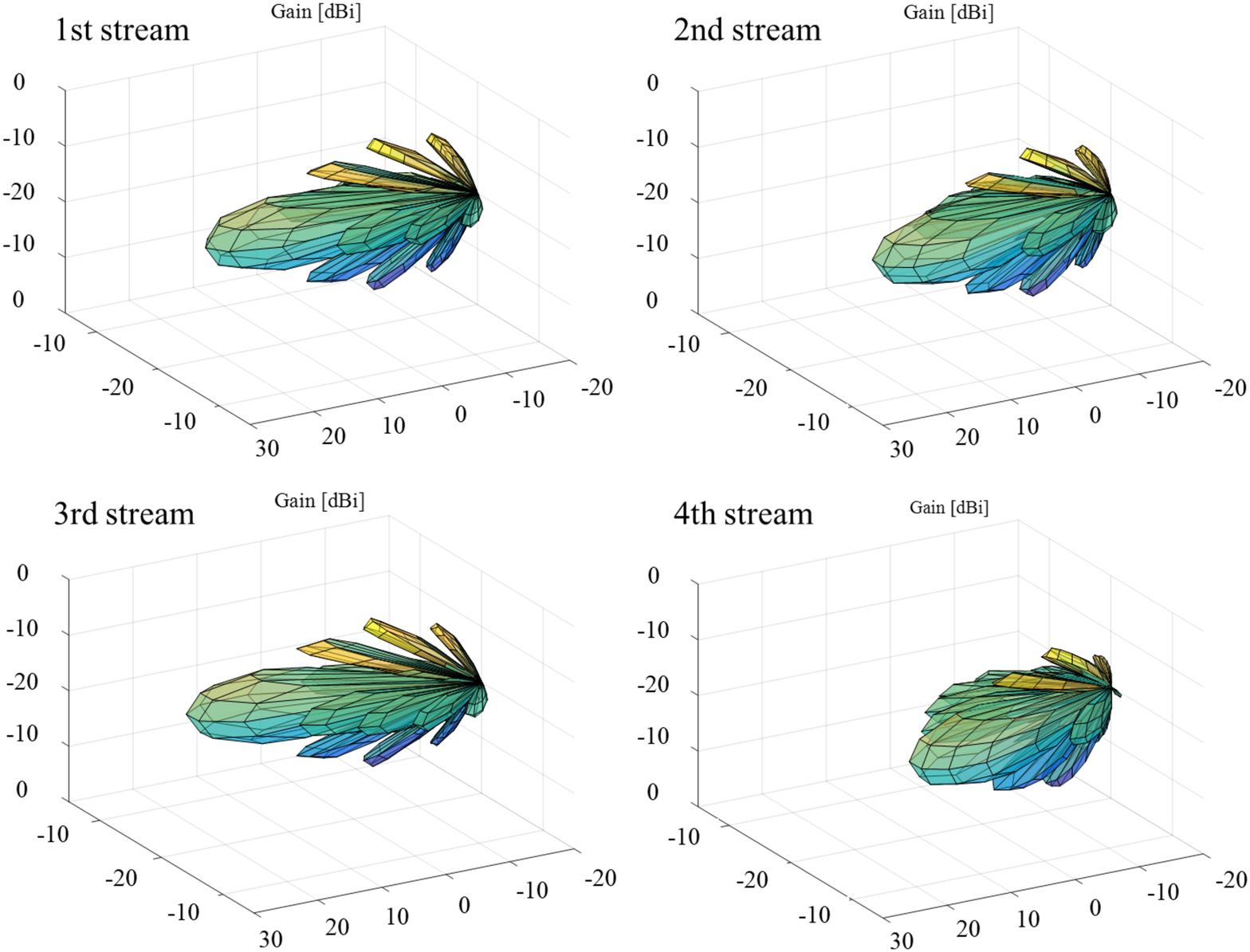}
    \caption{3-D beam patterns of Full-array derived by Determinant ($N_\mrm{UE}=4$).}
    \label{fig:3D_full2}
\end{figure}

\begin{figure}[htb]
    \centering
    \includegraphics[width=0.45\textwidth]{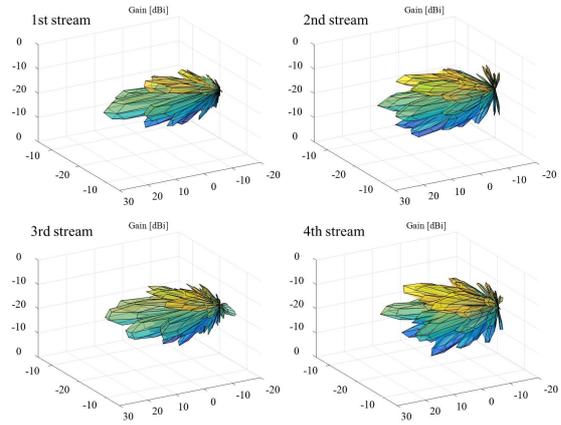}
    \caption{3-D semi-optimal beam patterns of Plane.}
    \label{fig:3D_Plane}
\end{figure}

\begin{figure}[htb]
    \centering
    \includegraphics[width=0.45\textwidth]{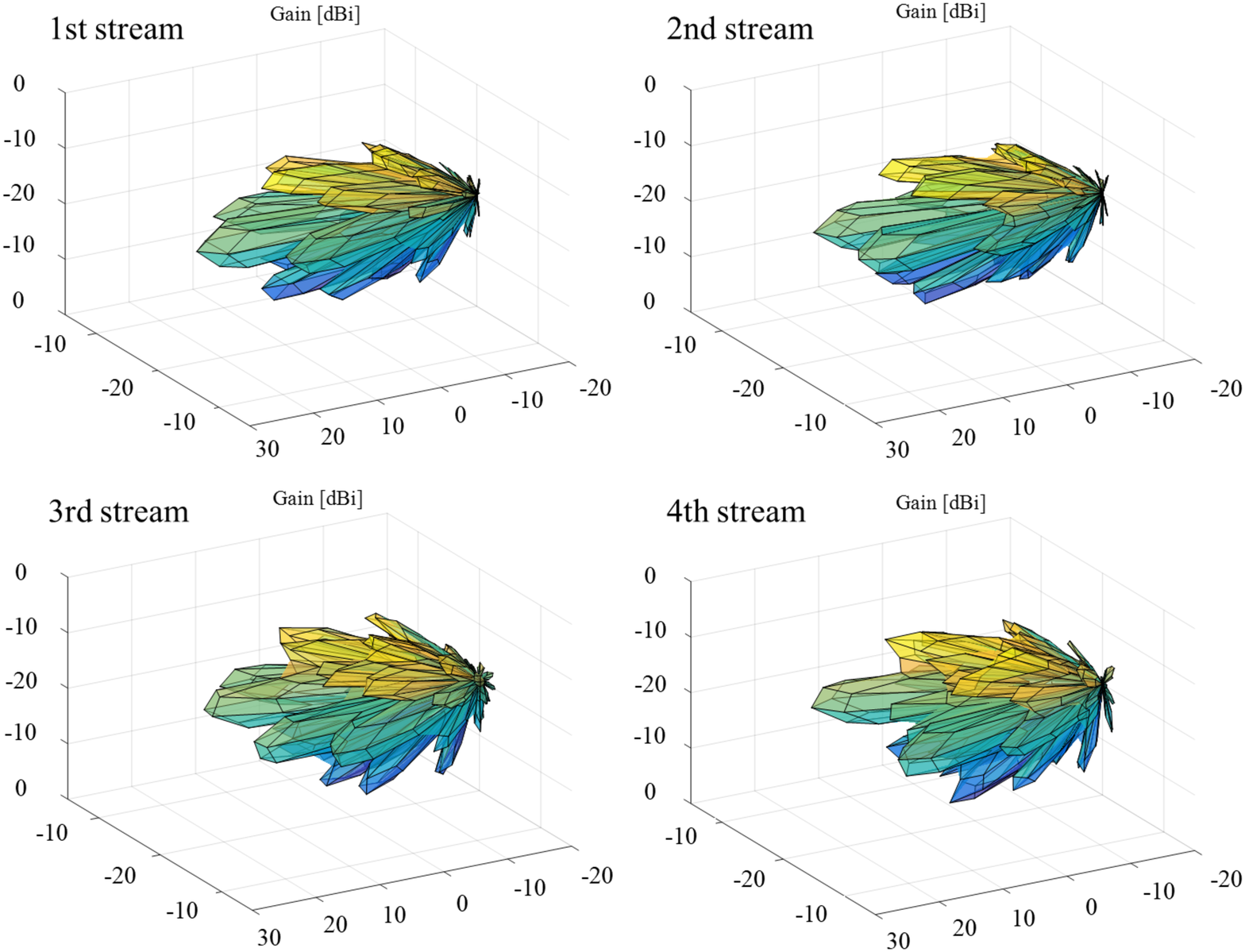}
    \caption{3-D semi-optimal beam patterns of Hemisphere.}
    \label{fig:3D_Hemi}
\end{figure}

To compare the proposed semi-optimal patterns to the conventional beam patterns, 
three cases of 3-D beam patterns are shown in Figs.\ \ref{fig:3D_full2} -  \ref{fig:3D_Hemi}. 
In the case of the conventional beam selection, one main lobe is mainly used
as shown in Fig.\ \ref{fig:3D_full2}. 
On the other hand, in the case of the beams of 
Plane (Fig.\ \ref{fig:3D_Plane}) and Hemisphere (Fig.\ \ref{fig:3D_Hemi}), 
it is found that there are multiple narrow beams in the range of the angular profile for all streams. 
It means that both the main and side lobes are useful by using OBPB method
and the received power becomes large in each stream. 
Thus, the channel capacity is improved.


\subsection{Channel correlation matrix with optimal beam patterns}
The normalized channel correlation matrix 
is derived which components are derived as 
\begin{align}
	|\tilde{r}_{i,j}| = \frac{|r_{i,j}|^2}{\sqrt{|r_{i,i}||r_{j,j}|}},
\end{align}
where $r_{i,j}$ is the $i$-th row and $j$-th column component of the channel correlation matrix $\mbf{R}_\mrm{BS,h}$ containing the derived beam patterns. 
The components and determinant of the channel correlation matrix in the case of $N_\mrm{UE}=4$
are shown 
in Table\ \ref{tbl:ChannelCorrelation}.

In the case of Hemisphere, 
the same number of streams and channel capacity with the optimal beam patterns 
are achieved. 
It is also found that 
the channel correlation matrix is completely orthogonalized
and the determinant is larger than that of the other methods. 
In the cases of 1/32 sphere and Plane, 
the channel correlation matrices are almost orthogonalized
and their determinants are much larger than that of the conventional hybrid beamforming. 
From the results,
it is found that using the beam patterns derived by OBPB improves the SU-MIMO system performance.

In the case of Full-array, 
the determinant of the channel correlation matrix 
by the beam selection using the determinant is larger than 
that using the received power because of low correlation. 
Therefore, the average channel capacity is slightly improved. 
In the case of Sub-array, the correlation coefficients between streams 
are lower than those of Full-array using the received power 
because the effective sources of the beam patterns are separated 
from each other
in Sub-array. 
It is found that Sub-array is preferable to Full-array using the beam selection 
in terms of the received power
when the correlation between selected beam patterns are high. 

\begin{table*}[tb]
	\begin{center}
	\caption{Normalized channel correlation matrix from 1st to 4th streams ($N_\mrm{UE} = 4$).}
	\label{tbl:ChannelCorrelation}
		\begin{tabular}{|c||c|c|}
		\hline
		Method & Normalized channel correlation matrix & Determinant [dB] \\
		\hline
		\hline
		Hemisphere
		&
		$\left[ 
			\begin{array}{cccc}
			1 & 0 & 0 & 0 \\
			0 & 1 & 0 & 0 \\
			0 & 0 & 1 & 0 \\
			0 & 0 & 0 & 1\\
			\end{array} 
			\right]$
		& 168 \\
		\hline
		1/32-sphere
		& 
		$\left[ 
			\begin{array}{cccc}
			1 & 0.00010 & 0.00052 & 0.00033 \\
			0.00010 & 1 & 0.00028 & 0.00066 \\
			0.00052 & 0.00028 & 1 & 0.00084 \\
			0.00033 & 0.00066 & 0.00084 & 1\\
			\end{array} 
			\right]$
		& 168 \\
		\hline
		Plane
		& 
		$\left[ 
			\begin{array}{cccc}
			1 & 0 & 0.014 & 0 \\
			0 & 1 & 0 & 0.021 \\
			0.014 & 0 & 1 & 0 \\
			0 & 0.021 & 0 & 1 \\
			\end{array} 
			\right]$
		& 144 \\
		\hline
		Full-array (Received power)
		& 
		$\left[ 
			\begin{array}{cccc}
			1 & 0.92 & 0.92 & 0.71 \\
			0.92 & 1 & 0.71 & 0.41 \\
			0.92 & 0.71 & 1 & 0.92  \\
			0.71 & 0.41 & 0.92 & 1 \\
			\end{array} 
			\right]$
		& 55.5 \\
		\hline
		Full-array (Determinant)
		& 
		$\left[ 
			\begin{array}{cccc}
			1 & 0.10 & 0.10 & 0.23 \\
			0.10 & 1 & 0.13 & 0.34 \\
			0.10 & 0.13 & 1 & 0.21 \\
			0.23 & 0.34 & 0.21 & 1 \\
			\end{array} 
			\right]$
		& 102 \\
		\hline
		Sub-array 
		& 
		$\left[ 
			\begin{array}{cccc}
			1 & 0.31 & 0.0025 & 0 \\
			0.31 & 1 & 0.31 & 0.0025 \\
			0.0025 & 0.31 & 1 & 0.31 \\
			0 & 0.0025 & 0.31 & 1\\
			\end{array} 
			\right]$
		& 102 \\
		\hline
		\end{tabular}
	\end{center}
\end{table*}

\subsection{Optimal number of streams and average channel capacity}
\begin{figure}[htb]
    \centering
    \includegraphics[width=0.42\textwidth]{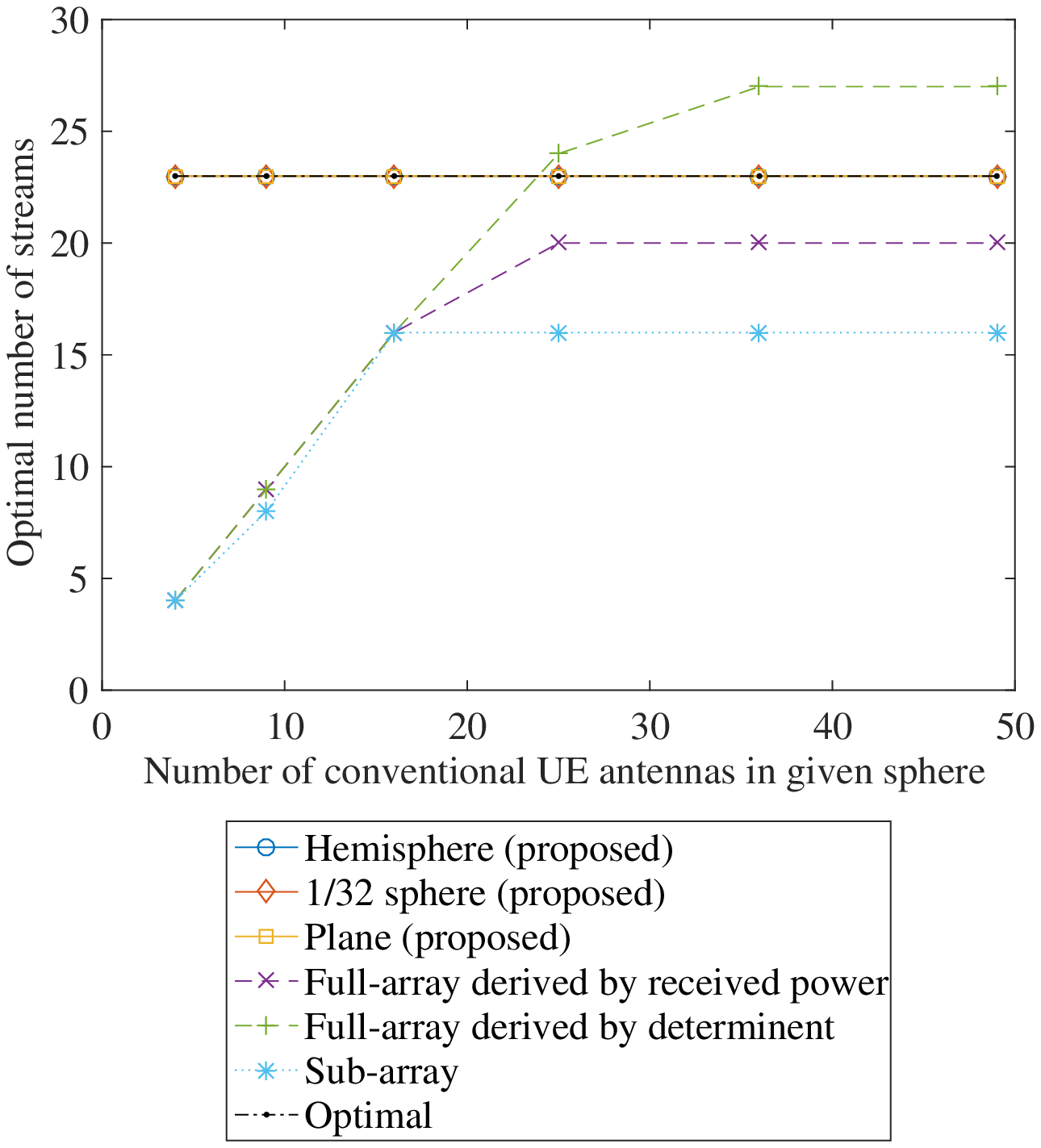}
    \caption{Optimal number of streams.}
    \label{fig:Stream}
\end{figure}

\begin{figure}[htb]
    \centering
    \includegraphics[width=0.42\textwidth]{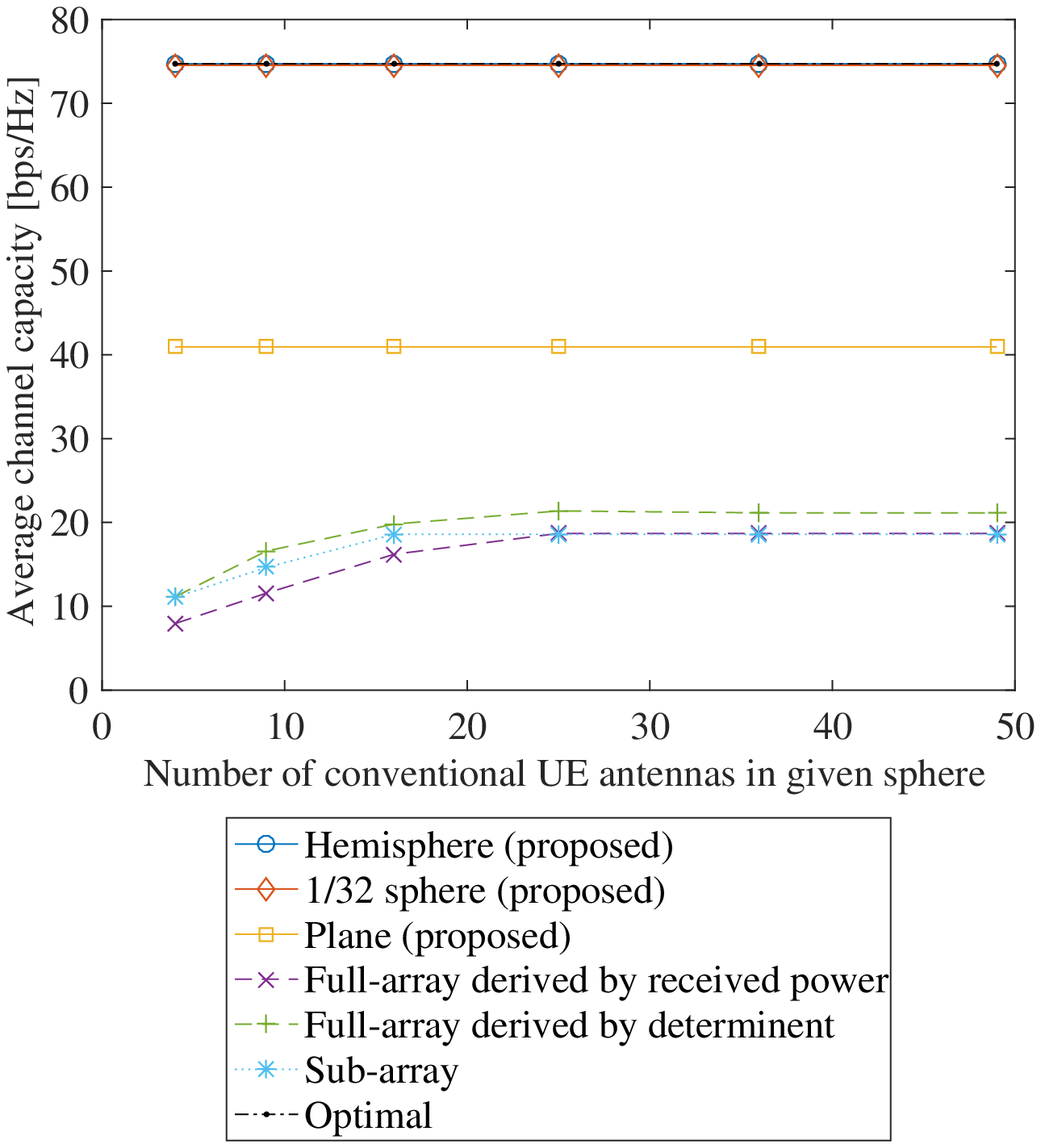}
    \caption{Average channel capacity.}
    \label{fig:Capacity}
\end{figure}

The optimal number of streams and the average channel capacity, 
corresponding to the number of conventional UE antennas in the given sphere, 
are 
depicted in Figs.\ \ref{fig:Stream} and \ref{fig:Capacity}. 
When the number of conventional UE antennas becomes large in Full-array, 
the optimal number of streams increases to 20 and 27
by using 
Received power and Determinant respectively. 
In Sub-array, the maximum value of the number of streams is limited to $N_\mrm{BS}/N_\mrm{BS,sub}$ which is up to 16 in the analysis. 
It is found that the average channel capacity also 
increases
and Sub-array is more effective than Full-array using Power 
when there is 
sufficient orthogonality of streams. 

In the case of Proposed, 
the optimal number of streams and the average channel capacity do not vary 
since the radii of antenna volume at both BS and UE sides are constant. 
Both the number of streams and the average channel capacity
increase by using the optimal patterns derived using OBPB
because the patterns are more matched to the angular profile than those of Full-array and Sub-array.
The average channel capacity becomes 3.5 times or larger than using Full-array and Sub-array 
in the cases of 1/32-sphere and Hemisphere. 
It is because the patterns match the angular profile and low correlated by their orthogonality. 


\section{Conclusion}
\label{sec:Conclusion}
In this paper, 
we proposed a method that can derive optimal beam patterns
of analog beamforming for SU massive MIMO 
by iterative optimization. 
We also derived
the semi-optimal beam patterns 
on the assumed antenna surface, such as Plane, 1/32-sphere and Hemisphere, 
by using OBPB. 
Numerical analyses showed that the proposal could achieve the same number of streams and channel capacity as offered by optimal beam patterns for the case of a hemispherical surface. Also, 
it is clarified that the average channel capacity 
is 3.5 times or larger by using the semi-optimal beam patterns derived by OBPB  
than that by using the conventional hybrid beamforming. 
The semi-optimal beam patterns yield orthogonal streams because
the patterns are matched to the angular profile and low correlated with each other. 
Therefore, the analog 
beamforming
by OBPB is more effective for SU-massive MIMO than the conventional analog beamforming as it offers higher average channel capacity.

{\bf Maki Arai}
received the B.E. and M.E. degrees in electrical and electronic engineering
from the Tokyo Institute of Technology, in 2010 and 2012. 
She joined the NTT Network Innovation Laboratories, Nippon Telegraph and Telephone Corporation (NTT) in 2012.
From 2015, she is a doctor course student at Tokyo Institute of Technology. 
Her current research interests are high speed wireless communication systems and analysis and design of MIMO antennas.
She received the Research Encouragement Award of the Institute of Electrical Engineers of Japan (IEEJ) in 2010, 
the Antenna and Propagation Research Commission Student Award from the Institute of Electronics, Information and Communication Engineers (IEICE) in 2012, 
and the Young Engineers Award from the IEICE in 2015.

{\bf Kei Sakaguchi}received the B.E. degree in electrical and computer
engineering from Nagoya Institute of Technology, Japan in 1996, and
the M.E. degree in information processing from Tokyo Institute of
Technology, Japan in 1998, and the Ph.D. degree in electrical and
electronic engineering from Tokyo Institute of technology in 2006.
Currently, he is a Professor at Tokyo Institute of Technology in Japan 
and at the same time he is working at Fraunhofer HHI in Germany as a Senior Scientist.
He received the Outstanding Paper Awards from SDR Forum and IEICE in 2004 and 2005, respectively, 
the Tutorial Paper Award from IEICE Communication Society in 2006, 
and the Bast Paper Awards from IEICE Communication Society in 2012, 2013, and 2015. 
He is currently playing a role of the Industry Panel co-chair in IEEE Globecom 2017.
His current research interests are 5G cellular networks, sensor networks, and wireless energy transmission. 
He is a member of IEEE.

{\bf Kiyomichi Araki}received the Ph.D. degree 
from Tokyo Institute of Technology, Japan, in 1978. 
In In 1979-1980 and 1993-1994, he was a visiting research scholar at University of Texas, 
Austin and University of Illinois, Urbana, respectively. 
Since 1995 to 2014 he has been a Professor at Tokyo Institute of Technology, 
and now an Emeritus Professor. 
He has numerous journals and peer review publications in RF ferrite devices, 
RF circuit theory, electromagnetic field analysis, software defined radio, array signal processing, 
UWB technologies, wireless channel modeling, MIMO communication theory, 
digital RF circuit design, information security, and coding theory.

\end{document}